\documentclass[sigconf]{acmart}

\usepackage[T1]{fontenc}
\usepackage{balance}
\usepackage{pdfpages}
\usepackage{subcaption}
\usepackage{siunitx}
\usepackage{graphicx}
\usepackage{amsmath}

\graphicspath{{./figures/}}

\newcommand{\dc}{\Delta RC}
\newcommand{\dci}[1]{\dc_{#1}}
\newcommand{\vl}[1]{\texttt{VLONG}#1}

\newcommand{\us}{\si{\micro\second}}
\newcommand{\ms}{\si{\milli\second}}
\newcommand{\kbps}{\si{\kilo bp\second}}
\newcommand{\ks}{Kolmogorov-Smirnov}

\copyrightyear{2018}
\acmYear{2018}
\setcopyright{acmlicensed}
\acmConference[ASIA CCS '18]{2018 ACM Asia Conference on Computer and Communications Security}{June 4--8, 2018}{Incheon, Republic of Korea}
\acmBooktitle{ASIA CCS '18: 2018 ACM Asia Conference on Computer and Communications Security, June 4--8, 2018, Incheon, Republic of Korea}
\acmPrice{15.00}
\acmDOI{10.1145/3196494.3196518}
\acmISBN{978-1-4503-5576-6/18/06}

\begin{document}
\title[Information Leakage and Covert Communication Between FPGA Long Wires]{Leaky Wires: Information Leakage and Covert Communication Between FPGA Long Wires}

\author{Ilias Giechaskiel}
\affiliation{\institution{University of Oxford}}
\email{ilias.giechaskiel@cs.ox.ac.uk}

\author{Kasper B. Rasmussen}
\affiliation{\institution{University of Oxford}}
\email{kasper.rasmussen@cs.ox.ac.uk}

\author{Ken Eguro}
\affiliation{\institution{Microsoft Research}}
\email{eguro@microsoft.com}

\begin{abstract}
  Field-Programmable Gate Arrays (FPGAs) are integrated circuits that
  implement reconfigurable hardware. They are used in
  modern systems, creating specialized,
  highly-optimized integrated circuits without the need to design and
  manufacture dedicated chips.
  As the capacity of FPGAs grows, it is increasingly common for designers to incorporate
  implementations of algorithms and protocols from a range of
  third-party sources. The monolithic nature of FPGAs means that
  all on-chip circuits, including third party black-box designs,
  must share common on-chip infrastructure, such as routing resources.
  In this paper, we observe that a ``long'' routing wire carrying a logical 1
  reduces the propagation delay of other adjacent but unconnected long wires
  in the FPGA interconnect, thereby leaking information about its state.
  We exploit this effect and propose a communication channel that can be used for both covert
  transmissions between circuits, and for exfiltration of secrets from the chip.
  We show that the effect is measurable for
  both static and dynamic signals, and that it can be detected using very small on-board
  circuits. In our prototype, we are able to correctly infer the logical state of an adjacent
  long wire over 99\% of the time, even without error correction, and
  for signals that are maintained for as little as 82us. Using a Manchester encoding
  scheme, our channel bandwidth is as high as 6kbps.
  We characterize the channel in detail and show that it is
  measurable even when multiple competing circuits are present and
  can be replicated on different generations and families of Xilinx devices
  (Virtex 5, Virtex 6, and Artix 7). Finally, we propose
  countermeasures that can be deployed by systems and tools designers to reduce the
  impact of this information leakage.
\end{abstract}

\includepdf[noautoscale]{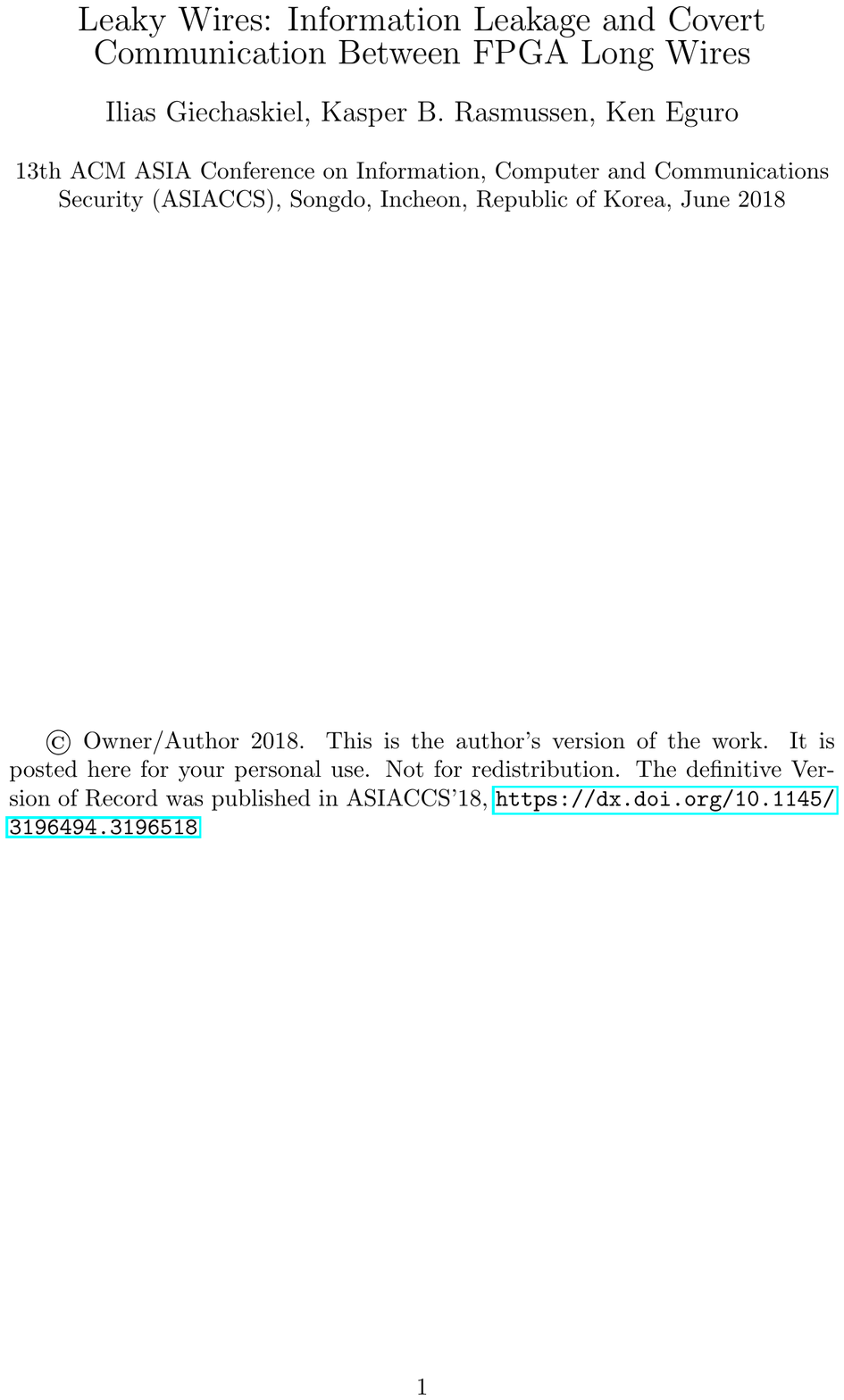}

\settopmatter{printacmref=false}
\renewcommand\footnotetextcopyrightpermission[1]{}
\maketitle

\section{Introduction}

The ever-increasing size and sophistication of FPGAs make them an
ideal platform for System-on-Chip integration.
FPGAs are often used in high-bandwidth, low-latency applications,
providing functionality such as network card replacement,
or massively parallel computation.
Besides permeating distributed systems and critical infrastructure,
FPGA chips are also integrated in end-products,
ranging from consumer electronics to medical and scientific equipment.
As a result, protecting their security is a necessary step to ensure that
their computations are performed in a trustworthy manner.

The high cost of design and development has led to an increase in
outsourcing, making it common to have designs
from different contractors on the same FPGA chip. Such
designs often include protocol and data structure implementations, or
more sophisticated circuits, like radio front-ends or soft
processors. This practice raises concerns about the malicious
inclusion of circuits (cores) that have additional backdoor
functionality. The cores can be functionally validated before being included in the
overall design, but such static analysis cannot always detect covert
channels~\cite{fpga_security}. It is therefore important to identify
and protect against such channels.

In this paper, we show that the value driven onto certain types of FPGA
routing resources, called ``long'' wires, influences the delay
of nearby wires, {\em even when the driven value remains constant}.
This distinguishes our approach from
prior work which depends on fast-changing signals~\cite{ro_crosstalk,asic_ht,ron},
and thus local voltage drops or inductive crosstalk.
Specifically, we find that if a long wire carries a logical 1, the
delay of nearby long lines will be slightly lower than when it carries
a logical 0. This difference in delay allows cores sharing the same
reconfigurable FPGA fabric to communicate, even when they are not directly
connected.

We demonstrate the phenomenon by building a transmitter and receiver,
which are unconnected, and only use adjacent long wires to communicate.
The receiver is a three-stage Ring Oscillator (RO), whose routing uses a long wire
between two of its stages. The transmitter drives a long wire adjacent
to that of the RO. When the transmitting wire carries a
logical 1, the routing delay of the RO long wire decreases, thereby
increasing the RO frequency. We detect these minor frequency changes
by counting the number of the RO signal transitions during a fixed time interval.
This mechanism can be used either for covert communication, or for the
exfiltration of fast-changing dynamic signals.

We conduct extensive experiments on three Xilinx FPGA families
and show that the phenomenon is independent of the device used, the
location and orientation of the transmitter and receiver,
and the pattern of transmission.  We perform all tests
on stock prototyping boards without modifications, 
and show that
the phenomenon can be detected even in the presence of environmental
noise and with only small circuits internal to the FPGA.
Finally, we propose new defense mechanisms which can be
implemented by systems and tools designers to reduce the
impact of this information leakage.

\section{Background}
\label{sec:background}

Field-Programmable Gate Arrays (FPGAs) are integrated circuits that
implement reconfigurable hardware. At a basic level,
they consist of blocks of configurable lookup tables (LUTs), which can be used to
represent the truth table of combinatorial functions. They also include registers to store
data, as well as programmable routing, which
determines how the LUTs and registers are interconnected. FPGAs can thus be used to represent all
computable functions, including emulating sophisticated circuits such as entire CPUs.

The Xilinx FPGAs used in our experiments internally have a grid layout,
whose fundamental building block is called a
{\em Configurable Logic Block} (CLB). It is composed of two {\em slices}, each of which contains
four LUTs and registers. Each CLB has an associated {\em switch matrix}, which contains
resources to connect elements within a CLB, and enables
CLBs to communicate with each other. There are multiple types of such communication wires,
which have different orientations and lengths. In this paper we focus on a
specific type of routing resource, called a {\em long}. Longs are a wire type
used to efficiently communicate between CLBs that are far apart, and can be {\em vertical}
(connecting elements with the same $x$ coordinate), or {\em horizontal} (same $y$ coordinate).
We have observed the phenomenon in both types of wires, but for brevity we limit our
discussion to vertical longs, or \vl{s}.
Due to the FPGA's routing topology, additional shorter wires are often
needed to connect certain elements via long wires. We will refer to these wires
as ``local routing''.

Usually, the details of how logic elements are placed and signals are routed
are transparent to the circuit designers. Designers define their desired logic, but the conversion
to a physical implementation is handled by the manufacturer tools.
Compiler directives for the manual routing of signals are available, but these are often only used
if the exact routing impacts functionality. In the absence of manual directives, the tools
may elect to use any wire,
including longs, to carry a given signal in the circuit, without alerting designers.

That said, user-designed circuits often share the FPGA with third-party implementations of
various protocols, data structures, and algorithms.
These licensed designs, called Intellectual Property (IP) cores or blocks, often come
in a pre-routed black-box format, to eliminate the variability of on-the-fly
routing and attain a known clock frequency.
As a result, the routing of these blocks is opaque to circuit designers, and
blocks created by different parties can use routing resources in the same
channel of long wires. As our paper shows, this use of nearby long wires
can enable malicious circuits to communicate covertly, or extract
information from other cores.

Ring Oscillators (ROs) are a type of circuit which consists of an odd number of
NOT gates, chained together in a ring formation (i.e., the output of the last
gate is fed back as input to the first gate). ROs form a bi-stable loop,
whose output oscillates between 1 and 0 (true and false). The frequency of oscillation
depends on the number of stages in the RO, the delay between the
stages, as well as voltage, temperature, and small variations
in the manufacturing process~\cite{ro_jitter}. ROs in FPGAs are used
as temperature monitors~\cite{ro_sensing},
True Random Number Generators (TRNGs)~\cite{trng}, and Physically-Unclonable
Functions (PUFs)s~\cite{puf_characterization}, while in this paper
we present a way to use them to detect the logic state of nearby wires.

\section{System and Adversary Model}
\label{sec:adversary}

\begin{figure}[tb]
  \centering
  \includegraphics[width=.75\linewidth]{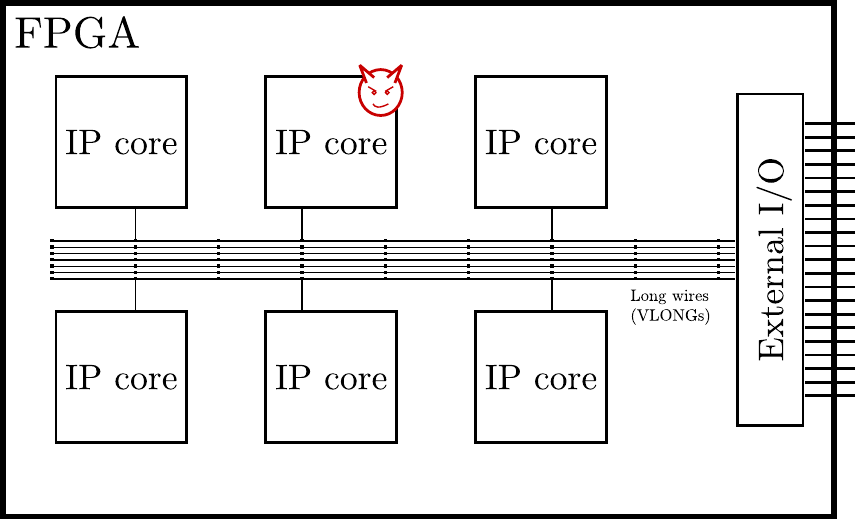}
  \caption{System model. Different IP cores share common FPGA routing
    resources. The cores can be supplied by different
    contractors and may contain malicious functionality.}
  \label{fig:system_model}
\end{figure}

FPGA designs contain IP cores sourced from
third-parties, and some of these cores may contain
unwanted functionality, as shown in Figure~\ref{fig:system_model}.
These third-party IP cores can be distributed as
fully-specified, pre-placed and pre-routed elements (``macros'') to meet timing constraints
(e.g., DDR controllers) and reduce compilation time,
with the macro repositioned at specific intervals where the logic and routing fabric is
self-similar~\cite{auto_macro,macro_clock,macros_compilation,hmflow}.

As FPGAs often process highly-sensitive information (e.g., cryptographic
keys), it is essential to ensure that data does not leak to unauthorized
third-parties. In this paper,
we focus on malicious IP cores which aim to infer information about the
state of nearby (but physically-unconnected) logic. The adversary
can thus insert one or more IP cores into the design, but these
cores are not directly connected. The adversary can also define the internal
placement and routing of his own
blocks and force his cores to use specific routing resources that can
compromise the integrity of a reverse-engineered target IP block.  Note that
directly connecting to the target IP block would result in a logical error in
the compilation flow, but merely using adjacent wires does not raise such
errors. We discuss how the adversary can accomplish his goals
in Section~\ref{sec:influencing}.

The adversary does not have physical access to the board, and
can thus not alter the environmental conditions or physically modify the FPGA board
in any way. There is also no temperature control
beyond the standard heatsink and fan already mounted on the FPGA, and we do not add
any special voltage regulation, or shielding to the chip or the connected wires. Such modifications
reduce noise and improve the stability of
measurements~\cite{ro_ht_analysis,static_leakage,side_receivers,ro_sensing}, and
would thus make it easier for the adversary to achieve his goals.

In this paper, we show that by using long wires, an adversary can infer the nearby
state of blocks he does not control,
or establish covert communication
between two co-operating IP cores under his control, even in the presence of
power and temperature fluctuations. We provide further motivation and applications
of the capabilities offered by this new source of information leakage in Section~\ref{sec:examples}.

\subsection{Motivation}
\label{sec:examples}

With increased outsourcing, Hardware Trojans (HTs) have become a
common-place security threat for FPGAs~\cite{ht_bitstream,ht_taxonomy}.
Adversarial IP cores can thus eavesdrop on nearby cores and attempt to
extract information about their state. As designs are often
tested to detect HTs and other security threats~\cite{asic_ht,ro_ht_analysis,ron},
we thus assume that the IP cores provide legitimate
functionality that is needed by the user, and that they do not contain additional
logic which would make them easy to detect.
Indeed, the transmitter and receiver we present have dual use,
hiding their malicious functionality in their routing,
not their actual combinatorial and sequential logic. As a result,
unlike conventional backdoors, our IP cores would pass
timing/netlist/bitfile verification, since they do not require
additional gates, presenting a bigger challenge to designers.

Multi-user setups present further threats beyond
a malicious core eavesdropping on signals not under the adversary's control.
Intel Xeon and other CPUs with integrated FPGAs bring FPGAs
closer to a traditional server model, while FPGAs in
cloud environments (e.g., Amazon EC2 F1 instances) are also becoming increasingly available.
Although these are currently allocated on a per-user basis,
we can expect that they will eventually become sharable commodity
resources, since FPGAs already allow for partial reconfiguration,
and designs exist where different processors have access to and can
re-configure the same FPGA chip~\cite{fpga_patent}.

An additional threat arises when IP cores of different security guarantees
are integrated on the same design~\cite{fpga_security,temperature_covert,fides}.
For example, an adversary implementing
the FM radio core on a phone SoC would want to eavesdrop on the
Trusted Platform Module's (TPM) AES encryption operations to recover its key.
As sensitive cores are highly scrutinized, an adversary who has also
implemented the TPM would want to establish a covert channel to transfer
the key using an inconspicuous transmitter.

Finally, the same phenomenon can be exploited
to watermark circuits~\cite{side_watermarks,voltage_side},
or introduce a no-contact debugging mechanism, for instance to
detect stuck signals, without altering routing.

\subsection{Influencing Placement and Routing}
\label{sec:influencing}

A potential issue with pre-placed and pre-routed
IP cores is that they are specific to an FPGA generation
(but can be used in different devices within the same family).
As we show in Section~\ref{sec:length}, however, the phenomenon
we present persists across 3 generations of Xilinx chips. As a result,
an adversary can provide an IP generation wizard that provides
different routing for different families, and
dynamically choose the placement of the IP cores.
In fact, as we show in Section~\ref{sec:location}, the location of the actual logic and wires is
not important, so the adversary merely needs to ensure that the transmitter and the receiver
use long wires which are adjacent.

If the adversary only pre-routes but does not pre-place his cores, he can
still succeed, even if he leaves the absolute placement of his cores to
the routing tools. Assume the FPGA has $N$ long wires, the transmitted signal can be recovered
from $w$ nearby wires, the receiver uses $R$ longs, and the transmitter uses $T$ longs. Then,
the probability that at least one segment of the transmitter is adjacent to a segment
of the receiver is $(R+T-1)\cdot w /N$, assuming the tools place the two cores
at random. For the FPGA boards we have used,
$N\approx 8,500$ (equal to the number of CLBs) and $w=4$, so with $R=T=5$, an adversary
has a 0.42\% chance of success. Since tools
do not pick locations at random or spread the logic,
the probability of success is higher in practice.
The adversary can also increase this probability by accessing
relatively unique elements such as Block RAM (BRAM), DSP blocks, or
embedded processors on the FPGA fabric.
For example, the devices we used have less than 150 DSP slices and 300 BRAM blocks,
so accessing them reduces the number of possible placements for the attacker's cores.

A more powerful adversary can instead subvert the compilation tools themselves,
which is a common threat model for FPGAs~\cite{fpga_security,lut_trojan}.
Note that, as before, since the final netlist itself is often verified
post-synthesis and -routing, the adversary still does not desire to include
additional logic in the design, but just affect the routing/placement of
his malicious cores.
Finally, in co-located multi-user instances,
the adversary {\em is} the user, so he can always choose
the location of his own cores, without the need to rely on the above.

\section{Channel Overview}
\label{sec:overview}

\begin{figure}[tb]
\centering
\includegraphics[width=0.85\linewidth]{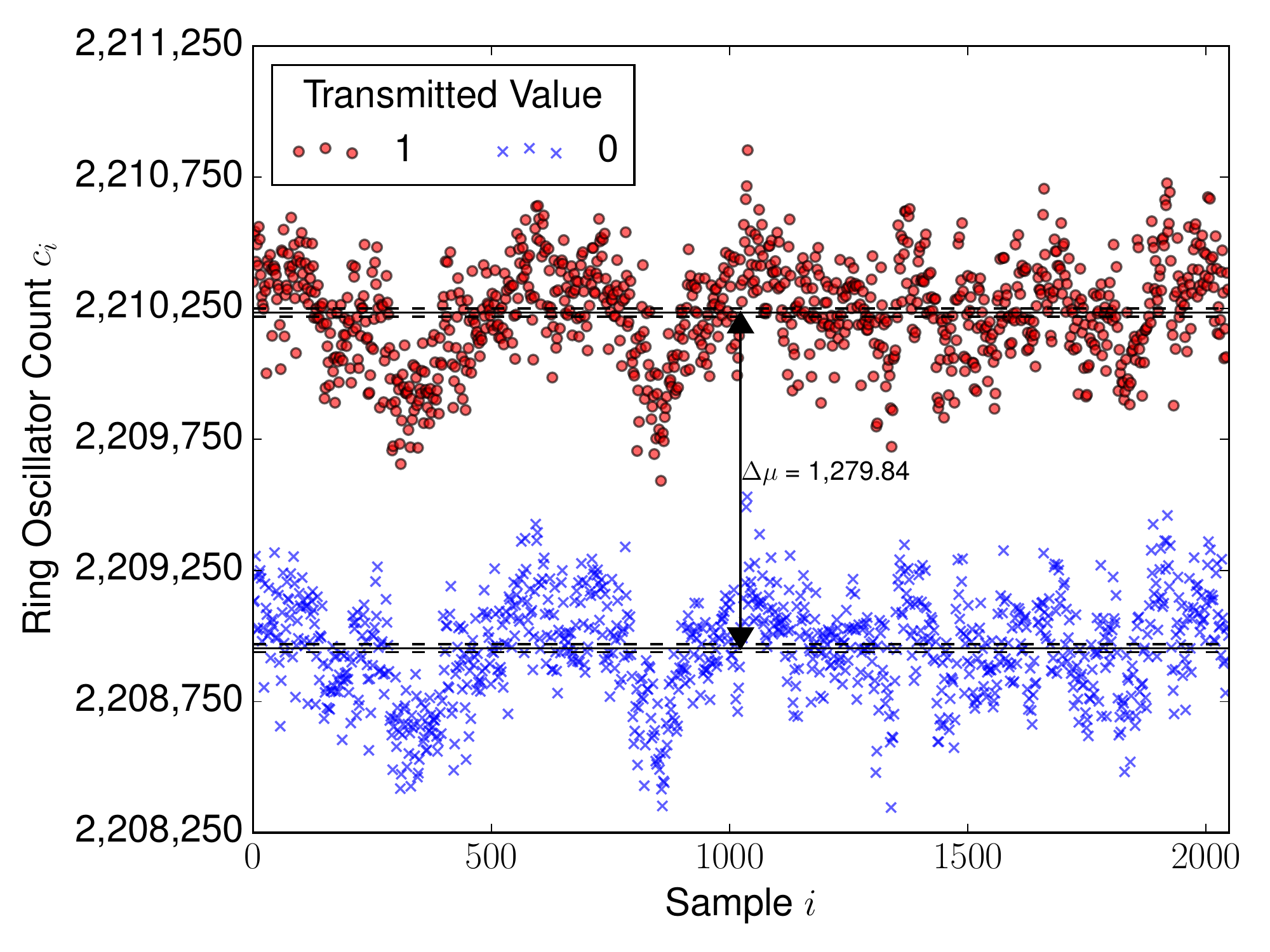}
\caption{Ring oscillator counts and 99\% confidence intervals for a setup where transmitter and
receiver use 5 longs each. The receiver is able to distinguish between signals using a
simple threshold, despite noise from the environment.}
\label{fig:v5_5}
\end{figure}

Our channel exploits the
fact that the delay of long wires depends on the
logical state of nearby wires, {\em even when the signals they are carrying are static}.
We find that when the
transmitting wire carries a~1, the delay of the nearby receiving wire
is lower, which results in a higher number of ring oscillator (RO) counts.
This is a distinct mechanism from prior research, which depends on the switching
activity of nearby circuits, which decreases RO frequency~\cite{asic_ht,ron},
as we also independently verify in Section~\ref{sec:local}.

This dependence on the logical state of the transmitter is shown in
Figure~\ref{fig:v5_5}. The red dots and blue x's are RO counts
when the transmitter wire carries a logical 1 or~0 respectively.  The
difference between the counts when transmitting 1s and 0s is clear,
even under local fluctuations due to environmental
and other conditions: even when the absolute frequencies of the
ring oscillator change, the difference between the two frequencies
remains the same.

In order to characterize the efficacy and quality of the communication
channel in detail,
we perform a number of
experiments, the setup of which is detailed in Section~\ref{sec:setup}.
We first show in Section~\ref{sec:patterns} that the strength of the effect does not depend on
the transmission pattern, by measuring the effect of an
alternating sequence of 0s and 1s, as well as that of long runs of
0s and~1s, and of pseudo-random bits. We illustrate that even
for fast-changing dynamic signals,
an eavesdropping attacker can obtain the fraction of 1s and~0s, i.e.,
the Hamming weight on the transmitting wire.

We then show that longer measurement
periods and overlaps make it easier to distinguish between different
bits in Section~\ref{sec:capacity}. The strength of the effect changes based on the receiver
and transmitter lengths, and this dependence exists across generations
of devices, but with a different magnitude. We also demonstrate that the
absolute location and orientation of the transmitter and receiver do not change
the magnitude of the effect in Section~\ref{sec:location}.

Finally, we show in Section~\ref{sec:others} that the channel remains strong, even if significant
computation is happening elsewhere on the device simultaneously, showing
that the channel can be used in a realistic environment.
We demonstrate that for the transmitted information to be detectable,
the transmitter and receiver wires need to be adjacent, but where
exactly and in what direction the overlap occurs is not
significant. This indicates that it may be difficult for designers to
protect themselves from eavesdropping, or detect intentional
malicious transmissions.

Overall, we show that the channel is stable across FPGA
generations, devices, and locations within a device. It is also
high-bandwidth, and can be used to implement both covert communications and
eavesdropping attacks, without tapping into existing signals, and with
minimal resources, as we explain in Section~\ref{sec:exploiting}.

\section{Experimental Setup}
\label{sec:setup}

\begin{figure}[tb]
\centering
\includegraphics[width=.85\linewidth]{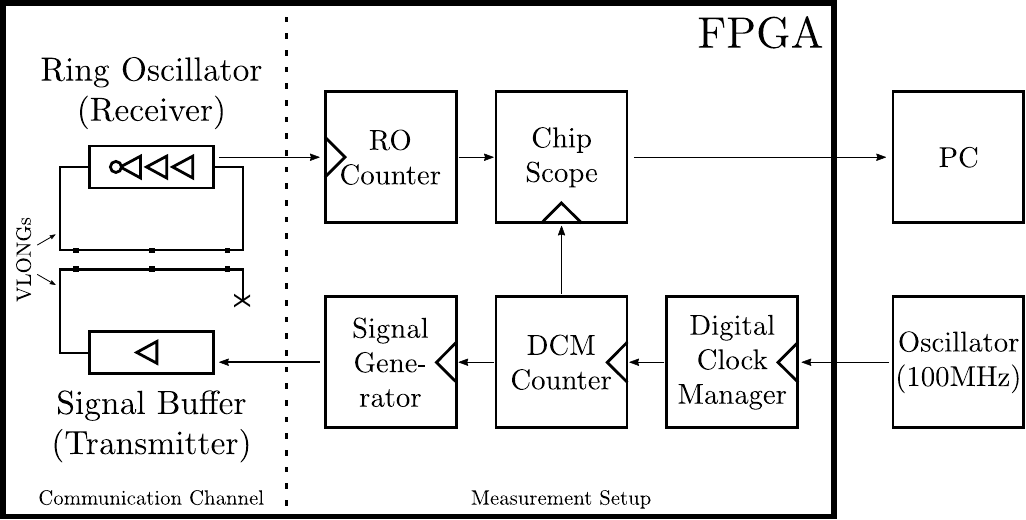}
\caption{Experimental setup. Transmitter and receiver use
long wires to communicate, while the measurement setup generates signals and
measures their effects.}
\label{fig:setup}
\end{figure}

In order to test the properties identified in Section~\ref{sec:overview},
we need to determine the factors we wish to vary, keeping the rest
of the setup fixed. This distinction naturally divides our experimental
setup into two parts, as shown in Figure~\ref{fig:setup}.
The communication channel circuit contains just the transmitter
and the Ring Oscillator receiver. The measurement half works independently of any specific
channel implementation,
generating the transmitted signal, sampling the RO counter,
and transferring the data to a PC for analysis.

The bulk of our experiments are conducted on three Virtex~5 XUPV5-LX110T (ML509) evaluations boards.
The boards include a heatsink and a fan, but we do not otherwise control for temperature,
and we also do not modify the board in any way (e.g., by bypassing the
voltage regulator) in accordance with our threat model.
Each experiment is run on every device 5 times, collecting 2048 data points per run, and
results are reported at the 99\% confidence level.

\subsection{Transmitter and Receiver}
\label{sec:cut_setup}

To illustrate the information leakage, our setup employs a minimal {\em transmitting} circuit:
the transmitter consists of a buffer LUT that drives one or more
long-wire segments connected end-to-end.
We use the term transmitter for brevity and because in the controlled experiments
we choose the value on the long wire, but the conclusions we draw are valid
whether transmissions are intentional or not.
The {\em receiving} circuit also uses long wires
that are adjacent to the transmitter's wire segments.
To measure the delay of the receiver's long wire segment(s),
we include it as part of a three-stage ring oscillator.
As in~\cite{side_receivers}, the oscillator contains one inverter (NOT gate) and two buffer stages.
The wire's delay directly influences the
frequency of oscillation, which we estimate by feeding the output of one of
the RO stages to a counter in our measurement setup.

The receiver and the transmitter are initially on fixed locations of the device, but we change
the location in Section~\ref{sec:location} to show that it does not influence our measurements.
We also change their lengths in Section~\ref{sec:length} to show that the effect becomes more
pronounced the longer the overlap is.

\subsection{Measurement Setup}
\label{sec:measurement_setup}

The measurement component generates the signals
to be transmitted and measures the RO frequency.
A new trigger event is produced every $N=2^n$ clock ticks. At every trigger,
the RO counter is read and reset, and a new value is presented to the transmitter.
For most experiments, the signal generator simply alternates between
0s and~1s, but we change the pattern in Section~\ref{sec:patterns} to show
the generality of the channel.

The 100MHz system clock
is driven by a Digital Clock Manager (DCM) to ensure clock quality.
For the majority of our experiments, we fix $n=21$ (corresponding to $2^{21}$
clock ticks, or $21\ms$), but vary $n$
in Section~\ref{sec:integration} to explore the accuracy vs. time trade-offs.
The sampled data is transferred to a PC for analysis through Xilinx's ChipScope
Integrated Logic Analyzer (ILA) core.

Unlike the circuit described above, the measurement logic is not hand-placed or hand-routed, due to
the large number of experiments performed.
Although the measurement logic could influence the RO frequency~\cite{ro_puf},
we repeat our experiments on multiple locations, control for other patterns, and average over
relatively lengthy periods of time.  Thus, we believe that any effects of the measurement circuitry
would influence the transmission of both zeros and ones equally,
a hypothesis we confirm in Section~\ref{sec:others}
by observing that the channel is only affected by adjacent wires.

\subsection{Relative Count Difference}
\label{sec:notation}

\begin{figure*}[tb]
  \centering\includegraphics[width=\linewidth]{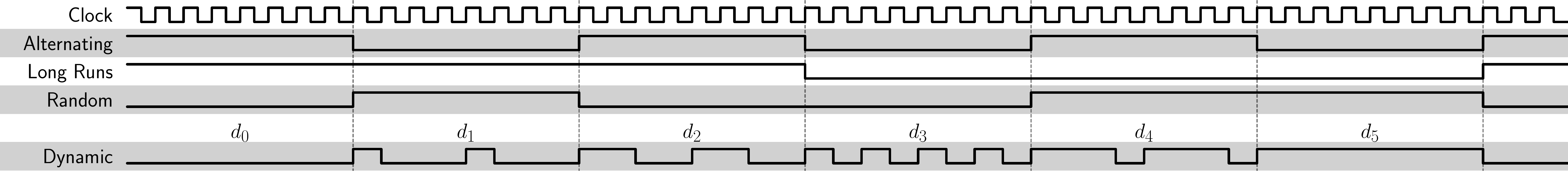}
  \caption{Timing diagram for the various transmission patterns used in the
           experiments. We test patterns which remain constant within a
           measurement period ($Alternating$, $Long\ Runs$, $Random$),
           and fast-changing patterns ($Dynamic$).}
  \label{fig:timing}
\end{figure*}

\begin{figure*}[t]
  \centering
  \resizebox{0.9\linewidth}{!}{%
    \subcaptionbox{Example run\label{fig:patterns_example}}{%
      \includegraphics[width=.43\linewidth]{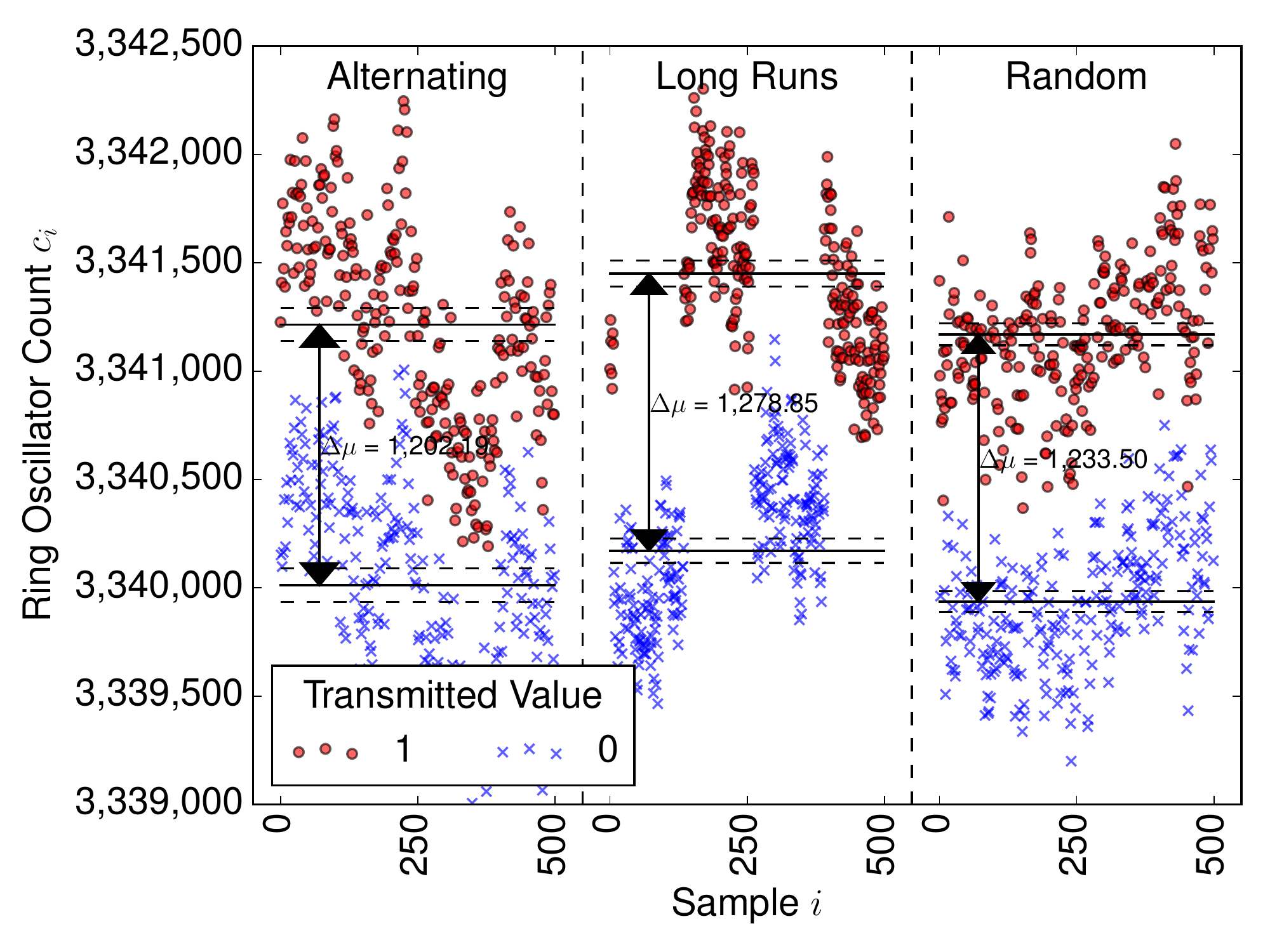}}\hspace*{4em}
    \subcaptionbox{Comparison across devices\label{fig:patterns_all}}{%
      \includegraphics[width=.43\linewidth]{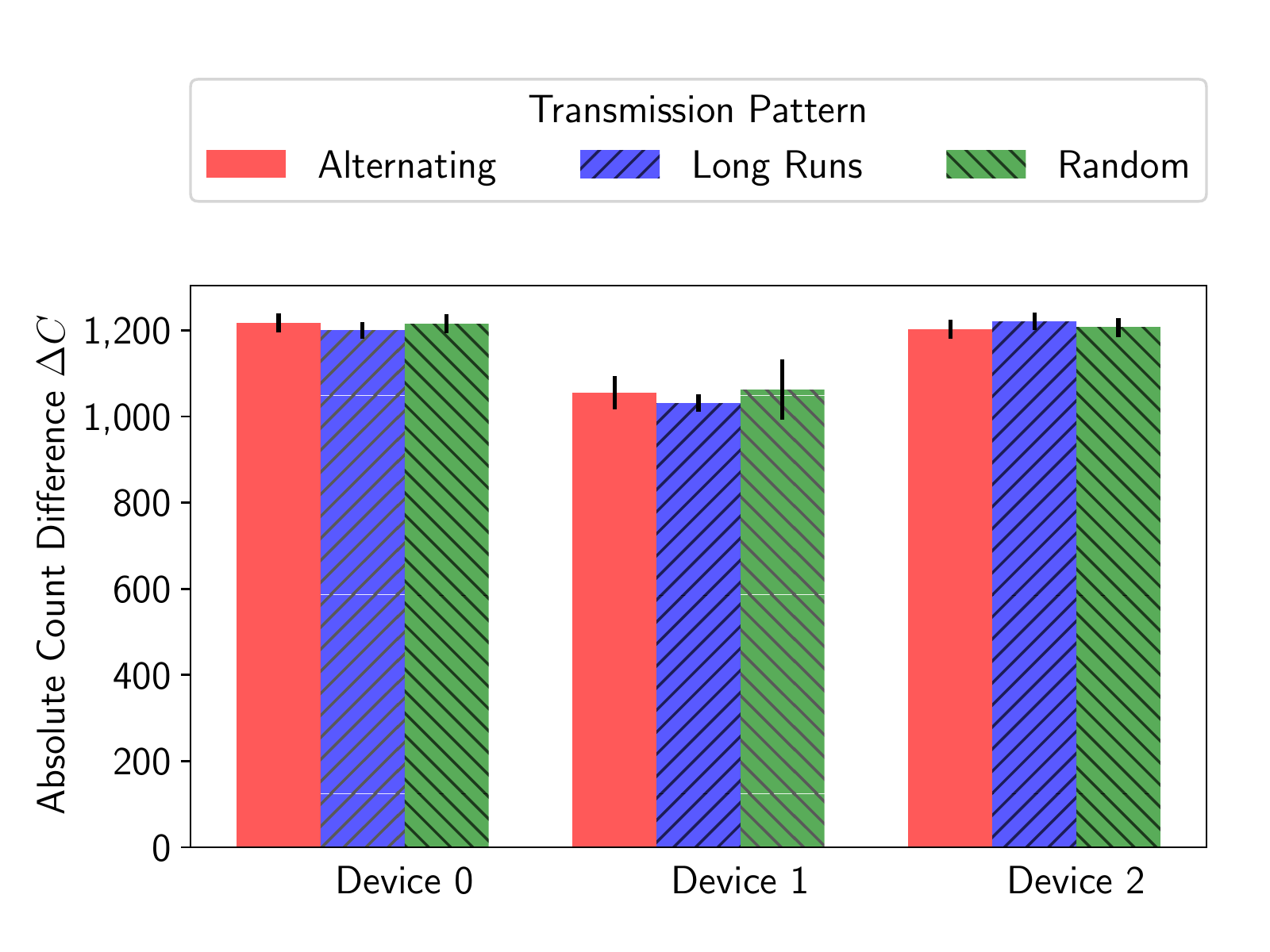}}}
  \caption{Effect of different static transmission patterns:
           (\subref{fig:patterns_example}) is a visualization of three
           different patterns: $Alternating$ (left), $Long\ Runs$ (middle),
           and $Random$ (right). (\subref{fig:patterns_all}) is a comparison
           across devices, with 99\% confidence intervals. The magnitude of
           the effect does not depend on the pattern used.}
  \label{fig:patterns}
\end{figure*}

When a clock of frequency $f_{CLK}$ is sampled every $m$ ticks and a ring oscillator
of frequency $f_{RO}$ driving a counter measures $c$ ticks, then
$f_{RO}/f_{CLK}\approx c/m$, with an appropriate quantization error
due to the unsynchronized nature of the RO and the system clock.  Thus,
\begin{equation}
\frac{f_{RO}^1-f_{RO}^0}{f_{RO}^1}\approx \frac{C^1 - C^0}{C^1}
\label{eq:df}
\end{equation}
where $C^i$ and $f^i$ represent the count and respective frequency when the
transmitter has value $i$. As a result, the relative change of frequency
can be approximated just by the measured counts, irrespective of the
measurement and clock periods.

In the basic setup, the transmitter alternates between sending zeros
and ones. We denote the $i$-th sampled count as $c_i$, so
the pair $p_i = (c_i, c_{i-1})$
always corresponds to different transmitted values.
For the sake of notation clarity, we will
assume that $c_{2i + 1}$ corresponds to a transmitted $1$
and we will be using the quantity
\[
\dci{i}=\frac{c_{2i+1}-c_{2i}}{c_{2i+1}}
\]
to indicate the relative frequency change between a transmitted
one and zero.  $\dc$ will denote the average of $\dci{i}$ over all measurement pairs $i$.
We discuss different transmission patterns
in Section~\ref{sec:patterns} and how to exploit the measurements in
Section~\ref{sec:exploiting}.

\section{Transmitter Patterns}
\label{sec:patterns}

In this section we show that the phenomenon observed does not
fundamentally depend upon the pattern of transmissions, i.e., that
only the values carried by the wire during the period of measurement matter,
and not the values that precede or follow it. We first show this
for relatively constant signals (Section~\ref{sec:static}), and then for highly dynamic
ones (Section~\ref{sec:dynamic}). Finally, we compare our results
to those produced by switching activity, which is traditionally discussed in
the context of Hardware Trojan detection (Section~\ref{sec:local}).

\subsection{Constant Signals}
\label{sec:static}

In the default setup, we use a slowly alternating signal, where
the transmitted value changes every sampling period. This pattern
is denoted by $Alternating$ in Figure~\ref{fig:timing}. In this experiment,
the transmitted value still remains constant within a given measurement period, and we sample
the ring oscillator at the same default rate (every $21\ms$), but change how
the signal generator chooses the next value to be transmitted.
The first additional pattern we test greatly slows down the alternation
speed of the transmitted signal. This $Long\ Runs$ pattern
maintains the same value for 128 consecutive triggers---in essence, testing the effects
of long sequences of zeros and ones.
The second setup employs a Linear Feedback Shift Register,
which produces a pseudo-random pattern of zeros and ones, and
is denoted by $Random$ in Figure~\ref{fig:timing}.

The results of this test are shown in Figure~\ref{fig:patterns}, with a sample of the data
in Figure~\ref{fig:patterns_example}, and a comparison across devices
in Figure~\ref{fig:patterns_all}. The RO counts remain
significantly higher when transmitting a 1 versus a 0, and the
average count difference remains identical, with almost no variability among the patterns.
We deduce that the pattern of transmission has
no persistent effect on the delay of nearby wires, allowing the channel
to be used without having to ensure a balanced distribution of transmitted values.

\subsection{Dynamic Patterns}
\label{sec:dynamic}

To show that the dominating factor in the observed phenomenon
is the duration for which the transmitter remains at a logical~1, and
{\em not} the switching activity of the circuit, we try various dynamic patterns.
As a result, even if a signal is not sufficiently long-lived,
the attacker can still deduce the signal's Hamming Weight (HW), and thus eavesdrop
on signals he does not control. We explain in Section~\ref{sec:exfiltration} how
to use this property to recover cryptographic keys through repeated measurements.

The dynamic patterns used are denoted by $Dynamic$ in the timing diagram of Figure~\ref{fig:timing}.
During each sampling period,
we loop the transmitter quickly through a 4-bit pattern at 100MHz.
We test six different 4-bit patterns,
only updating the looped pattern at each new sampling period.  For example,
for the pattern 1100 ($d_2$ in Figure~\ref{fig:timing}),
the transmitter would stay high for two 100 MHz clock ticks,
then low for two clock ticks, then back to high for 2 ticks, etc., until the end of the
sampling period. The six 4-bit patterns used are: $d_0=0000$, $d_1=1000$, $d_2=1100$, $d_3=1010$,
$d_4=1110$, and $d_5=1111$. These patterns respectively have a HW of
$0$, $25$, $50$, $50$, $75$, and $100$\%, while their switching frequencies are
$0$, $f=f_{CLK}/8$, $f$, $2f$, $f$, and $0$ respectively.

\begin{figure}[tb]
  \centering
  \includegraphics[width=0.85\linewidth]{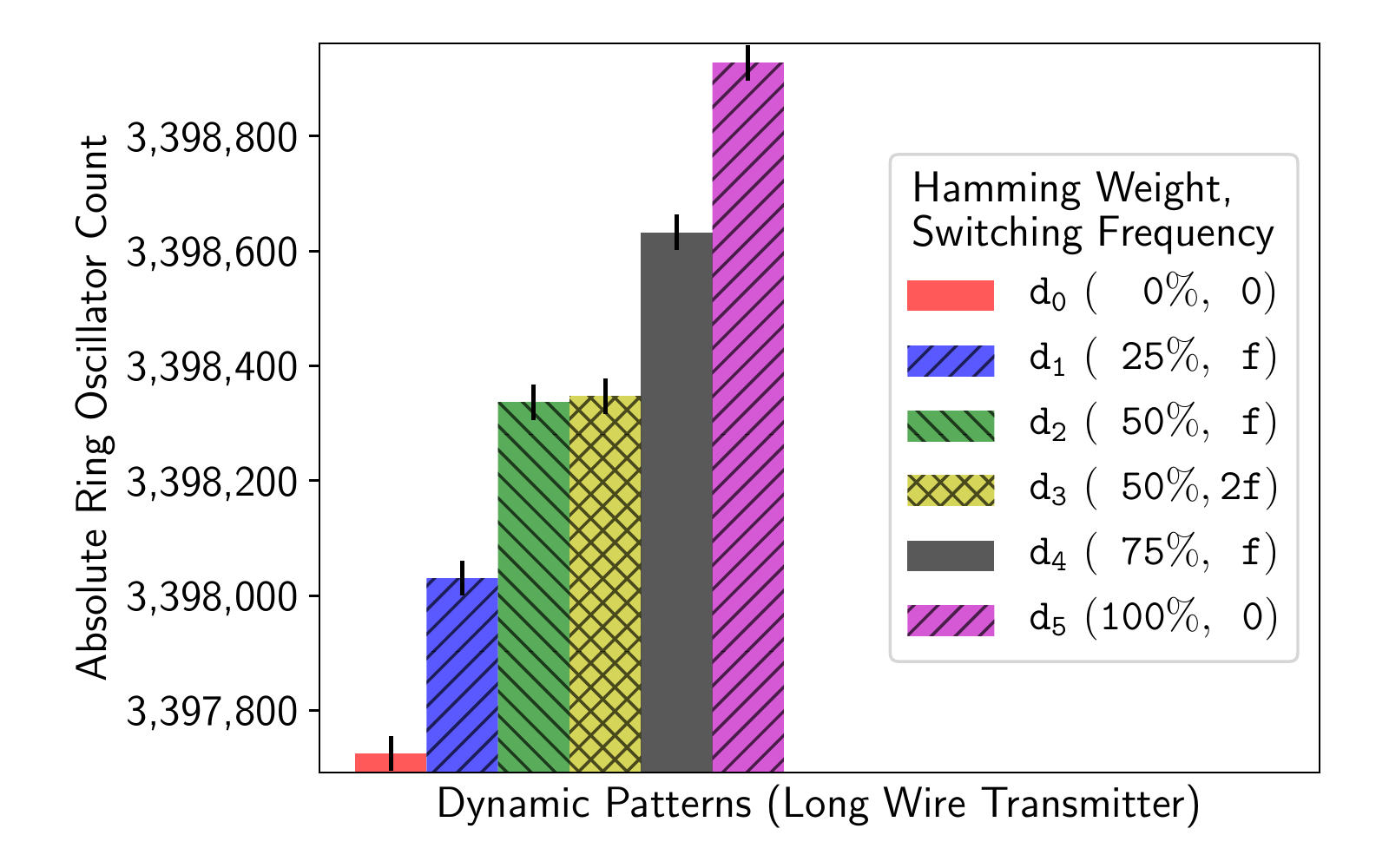}
  \caption{Effect of dynamic switching activity using a long-wire transmitter.
  RO counts increase with the Hamming Weight, but not with the switching frequency.}
  \label{fig:dynamic}
\end{figure}

Figure~\ref{fig:dynamic} shows the average count $C_i$ of the ring oscillator for each
of pattern $d_i$. We see that the RO frequency increases with the Hamming Weight,
so that $C_0<C_1<C_2\approx C_3 < C_4 < C_5$. However, the frequency is
otherwise unaffected by the switching transmission activity:
the \ks{} test suggests that there is no statistically significant difference between
the two distributions for $d_2$ and $d_3$.
Note that the receiver would not able to distinguish between
patterns $d_2$ and $d_3$ as a result (or more generally, any
patterns with the same Hamming Weight),
but we explain how to overcome this
limitation in Section~\ref{sec:exfiltration}.

\subsection{Local Routing}
\label{sec:local}

\begin{figure}[tb]
  \centering
  \includegraphics[width=0.85\linewidth]{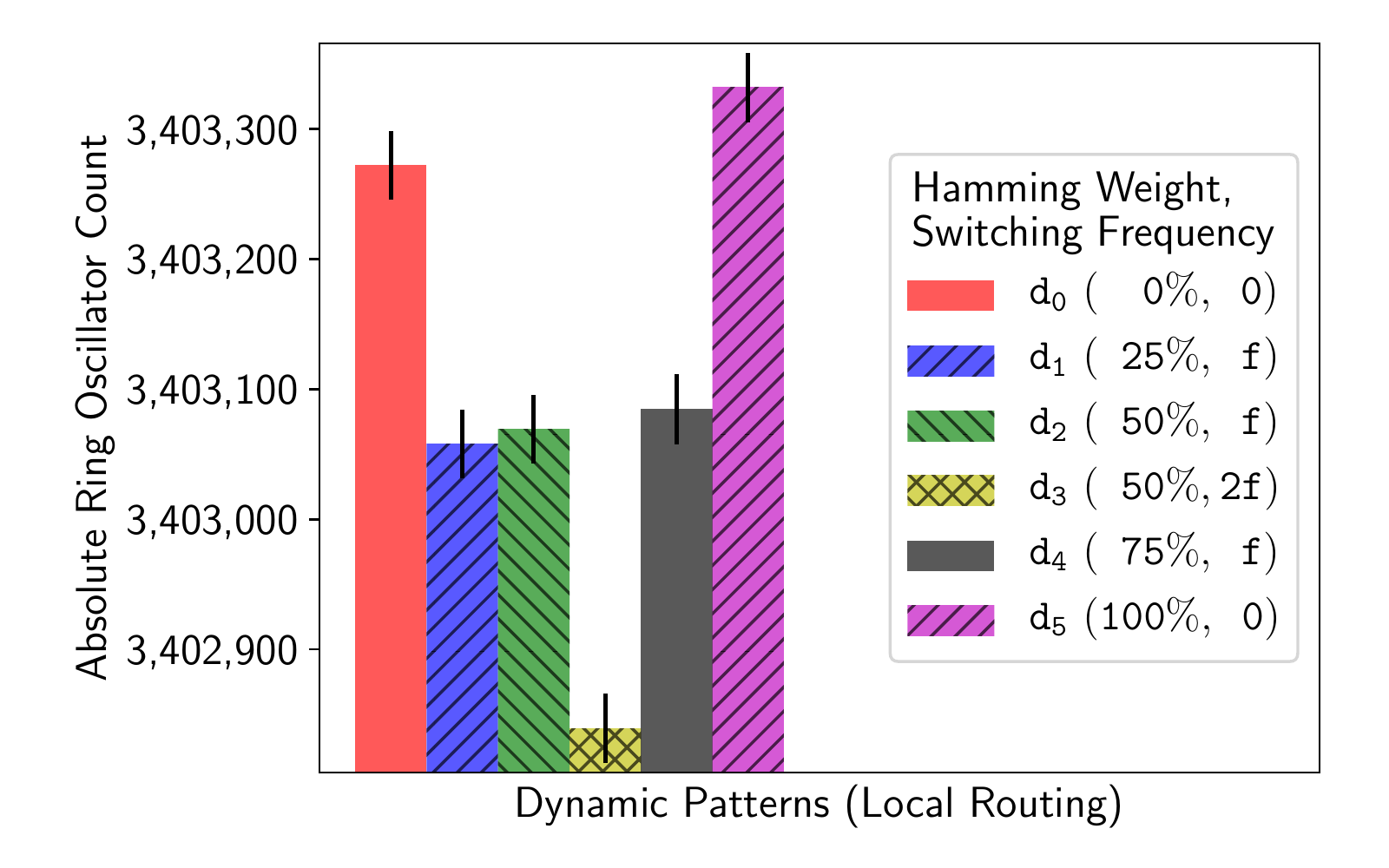}
  \caption{Effect of dynamic switching activity without long-wire overlaps.
  RO counts decrease with switching frequency, and are almost unaffected
  by the Hamming Weight.}
  \label{fig:dynamic_local}
\end{figure}

In this section, we show that when the two circuits do not have overlapping
long wires, switching activity {\em decreases} the oscillation frequency
of the RO. This reproduces the results reported by prior research
on Hardware Trojan detection~\cite{asic_ht,ron} and
also allows us to sanity-check our measurement setup.
To test this dependence on the long wire overlap, we remove the transmitter
using long wires, and replace it with a buffer of $312$ consecutive LUTs packed
into $39$ CLBs, using only local intra- and inter-CLB routing. We then
drive the same 6 dynamic patterns from Section~\ref{sec:dynamic} through the buffer,
and measure the results in Figure~\ref{fig:dynamic_local}. We can clearly see that
the ordering of the patterns exactly mirrors
their relative switching activity, with the RO counts $C_i$ corresponding to $d_i$
decreasing with increased
switching activity: $C_3 < C_1 \approx C_2 \approx C_4 < C_0 < C_5$.
The difference between the patterns with the same switching activity $d_1$, $d_2$,
$d_4$ is not significant according to the \ks{} test, but the count is slightly
higher for $d_5$ compared to $d_0$, which have no switching activity. This suggests
that the phenomenon we have identified and which reduces delay may be present for
shorter wires as well, but is considerably weaker, and requires much bigger circuits.
Overall, we can conclude that when the transmitter does
not use longs which overlap with the receiver and generates a lot of
switching activity through multiple redundant buffers, then
the observed RO frequency is indeed reduced, reproducing the results of prior work.

\section{Measurement Parameters}
\label{sec:capacity}

In this section, we discuss the trade-offs between the quality of the channel and the
measurement time (Section~\ref{sec:integration}), and length of overlap
between receiver and transmitter (Section~\ref{sec:length}).

\subsection{Measurement Time}
\label{sec:integration}

\begin{figure}[tb]
  \centering
  \includegraphics[width=0.85\linewidth]{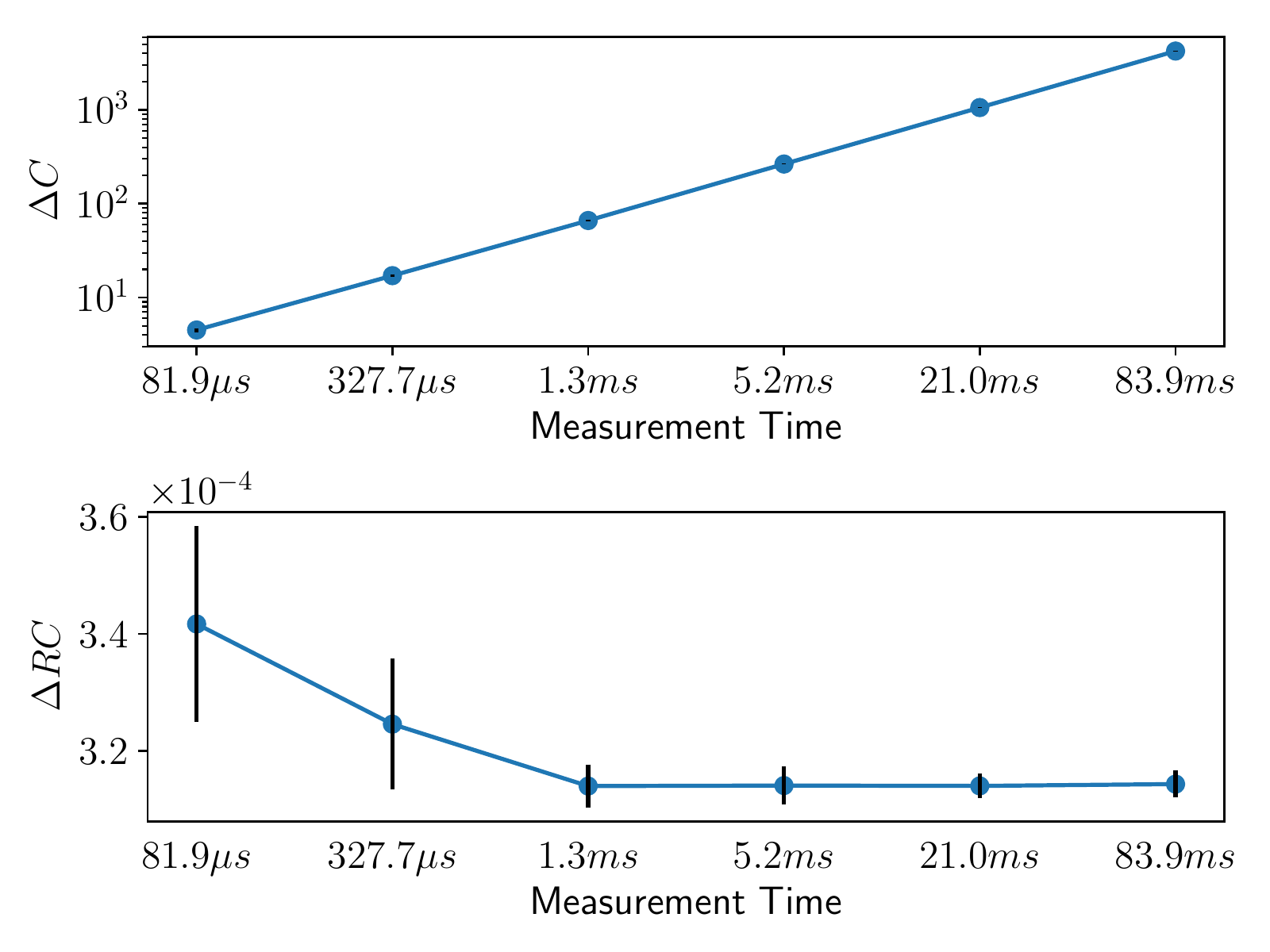}
  \caption{Absolute and relative count differences and~99\% confidence intervals
  for various measurement times. For a given transmitter and receiver overlap,
  the absolute magnitude of the effect increases 
  linearly with time.}
  \label{fig:integration}
\end{figure}

\begin{figure*}[tb]
  \centering
  \resizebox{\linewidth}{!}{
  \subcaptionbox{Virtex 5\label{fig:length_v5}}{%
    \includegraphics[width=.33\linewidth, trim=20 0 20 0, clip]{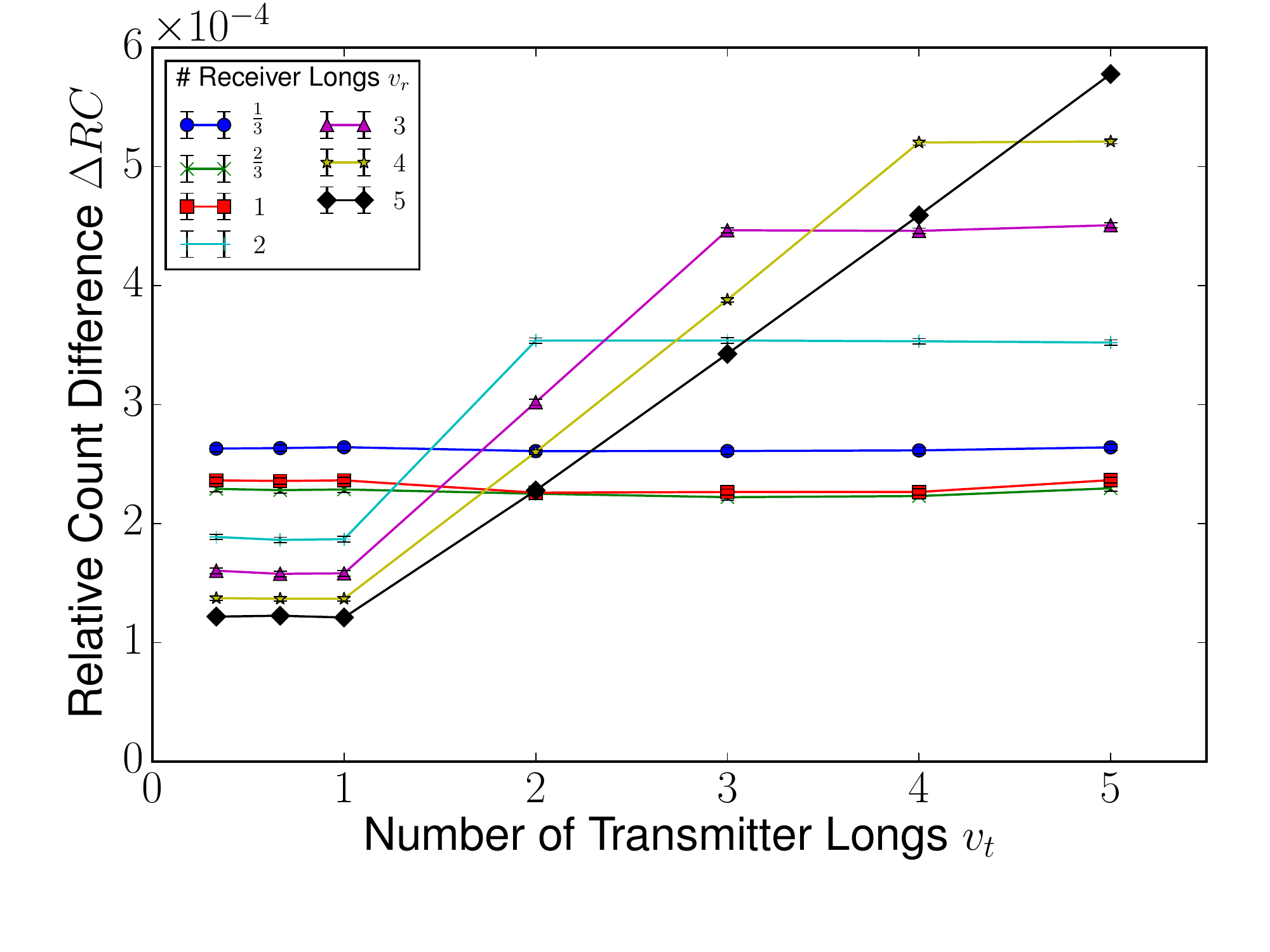}}\quad
  \subcaptionbox{Virtex 6\label{fig:length_v6}}{%
    \includegraphics[width=.33\linewidth, trim=20 0 20 0, clip]{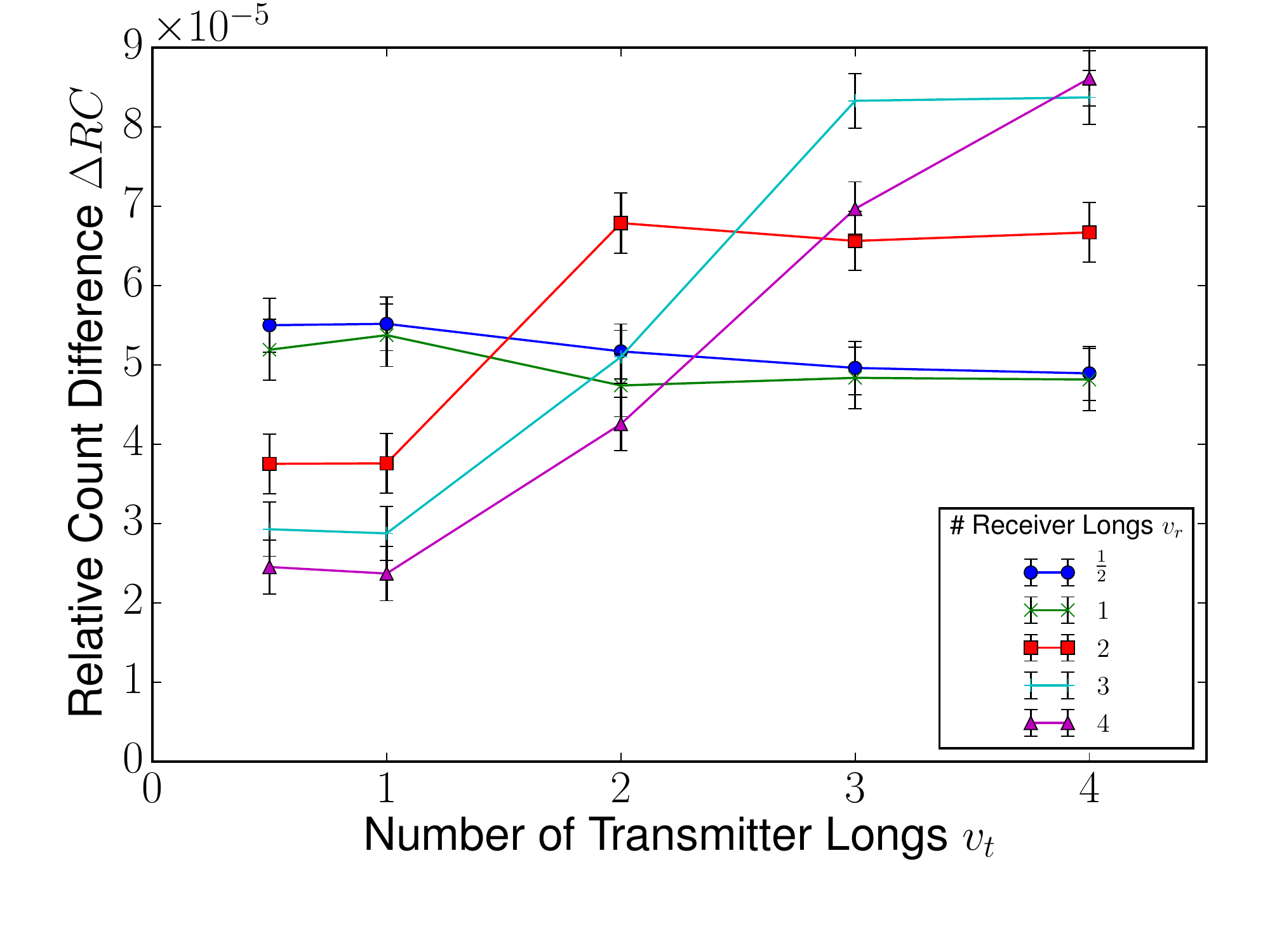}}\quad
  \subcaptionbox{Artix 7\label{fig:length_a7}}{%
    \includegraphics[width=.33\linewidth, trim=0 0 20 0, clip]{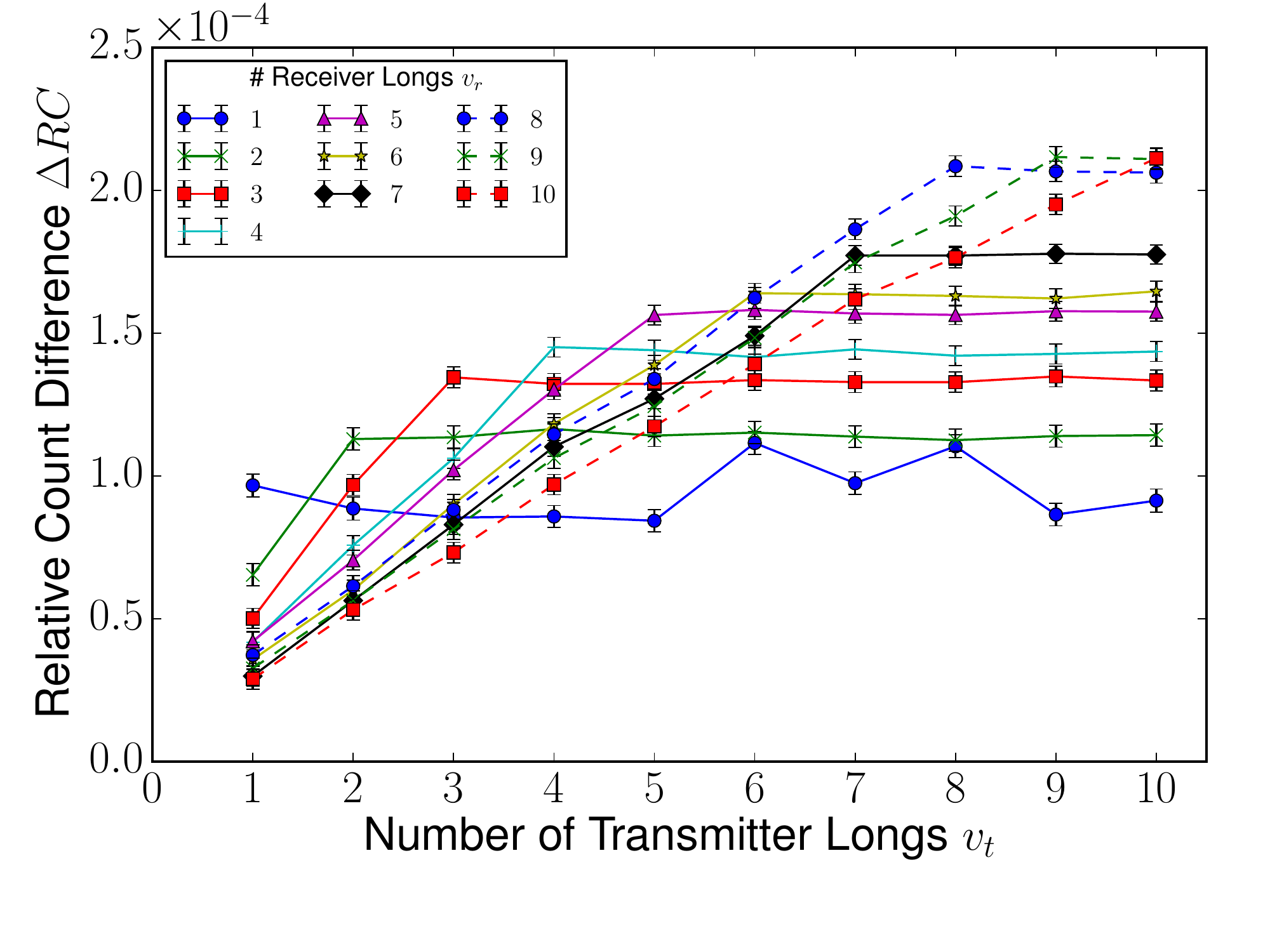}}}
\caption{Relative frequency changes $\dc$ with 99\% confidence intervals as a
         function of the transmitter and receiver lengths for different FPGA
         generations. The count difference is proportional
         to the overlap between the transmitter and the receiver.}
\label{fig:length}
\end{figure*}

In this experiment,
we return to the alternating pattern shown in Figure~\ref{fig:timing}, and vary the measurement time
by repeatedly quadrupling it. Both the absolute and the relative count difference
for the various times are shown in Figure~\ref{fig:integration}.
In the top of the figure, we see that
the absolute count difference $\Delta C$ grows linearly with increasing measurement
time. Hence, the RO count differences can be amplified, proportionally to the
duration of the measurement.

The relative differences $\dc$ (shown in the bottom of
Figure~\ref{fig:integration}) remain approximately constant for measurement periods above $1\ms$,
in accordance with our theoretical prediction of Equation~(\ref{eq:df}).
The values for shorter measurement periods are still close, but are far noisier:
for short measurement periods,
the absolute difference is small (${\approx}4$ for $t=82\us$), increasing quantization
errors, and making it harder to distinguish between signal and noise.
These results indicate that for a given receiver/transmitter placement,
the absolute magnitude of the effect depends solely on measurement time, with
longer measurement periods making it easier to distinguish between signals and noise.
An adversary can thus choose the measurement time, trading
throughput for lower bit error rate.

\subsection{Wire Length}
\label{sec:length}

We also characterize the effect of varying the length (number) of transmitter and receiver
wires $v_t$ and~$v_r$ in three generations of devices. Besides the Virtex~5 device we
have been using so far, we also measure the effect on a Virtex 6 ML605 and on an
Artix 7 Nexys 4.
The relative change in frequency $\dc$ is shown for different combinations
of $v_t$ and $v_r$ in Figure~\ref{fig:length}, for one device per generation. We notice
the same common pattern for all 3 generations of devices.

For a given number of long wires $v_r$ used by the ring oscillator,
there are 3 distinct segments for $\dc$
as the number of transmitter longs $v_t$ increases. The first segment occurs for
transmitters which use only parts of a long. Using partial wires is possible because even though
\vl{s} can only be driven from the top or the bottom, they have additional intermediate ``taps''
which can be used to read the values of the signal they carry.
In practice, using partial wires does not have an effect on the strength of the phenomenon:
$\dc$ remains constant for all fractions of a long. This result is to be expected since,
electrically, the entire long wire is driven even if the output tap does not take
full advantage of its length.

The second segment is the region where $v_t \le v_r$.  Here, $\dc$ increases
linearly with $v_t$, suggesting that the phenomenon
affects the delay of each long wire equally. The final region consists of
$v_t>v_r$, where $\dc$ remains constant.
The reason for this pattern is that there is no additional overlap between the newly
added segments of the transmitter and the receiver.

We also identify the effect of a given number of transmitter wires $v_t$
on receivers using a different number of longs $v_r$. Among receivers with $v_r \ge v_t$,
a smaller $v_r$ results in a larger effect. As an example, for $v_t=3$,
the effect for $v_r=5$ is smaller than it is for $v_r=3$. This behavior is due to the transmitter
affecting only the first $v_t$ out of $v_r$ long wire segments of the ring oscillator.
For smaller ring oscillators, these $v_t$ segments represent a larger portion of
the number of wires used, and hence of overall delay.

The opposite is true when $v_r \le v_t$: the larger the RO, the bigger
the resulting effect. For instance, for $v_t=4$, the effect for $v_r=3$ is
larger than the effect for $v_r=1$. This difference exists because even though the
delay of the routing scales linearly, the delay associated with the inverter and buffer LUT stages
remains constant.  Thus, the routing delay represents a larger fraction
of the overall delay (routing delay plus stage delay) for larger ROs.
Since this phenomenon only acts on routing delay,
larger ROs are affected more than shorter ones.

\section{Location Independence}
\label{sec:location}

\begin{figure*}[t]
  \centering
  \resizebox{\linewidth}{!}{
  \subcaptionbox{Absolute Location\label{fig:location}}{%
    \includegraphics[width=.33\linewidth, trim=20 0 40 0, clip]{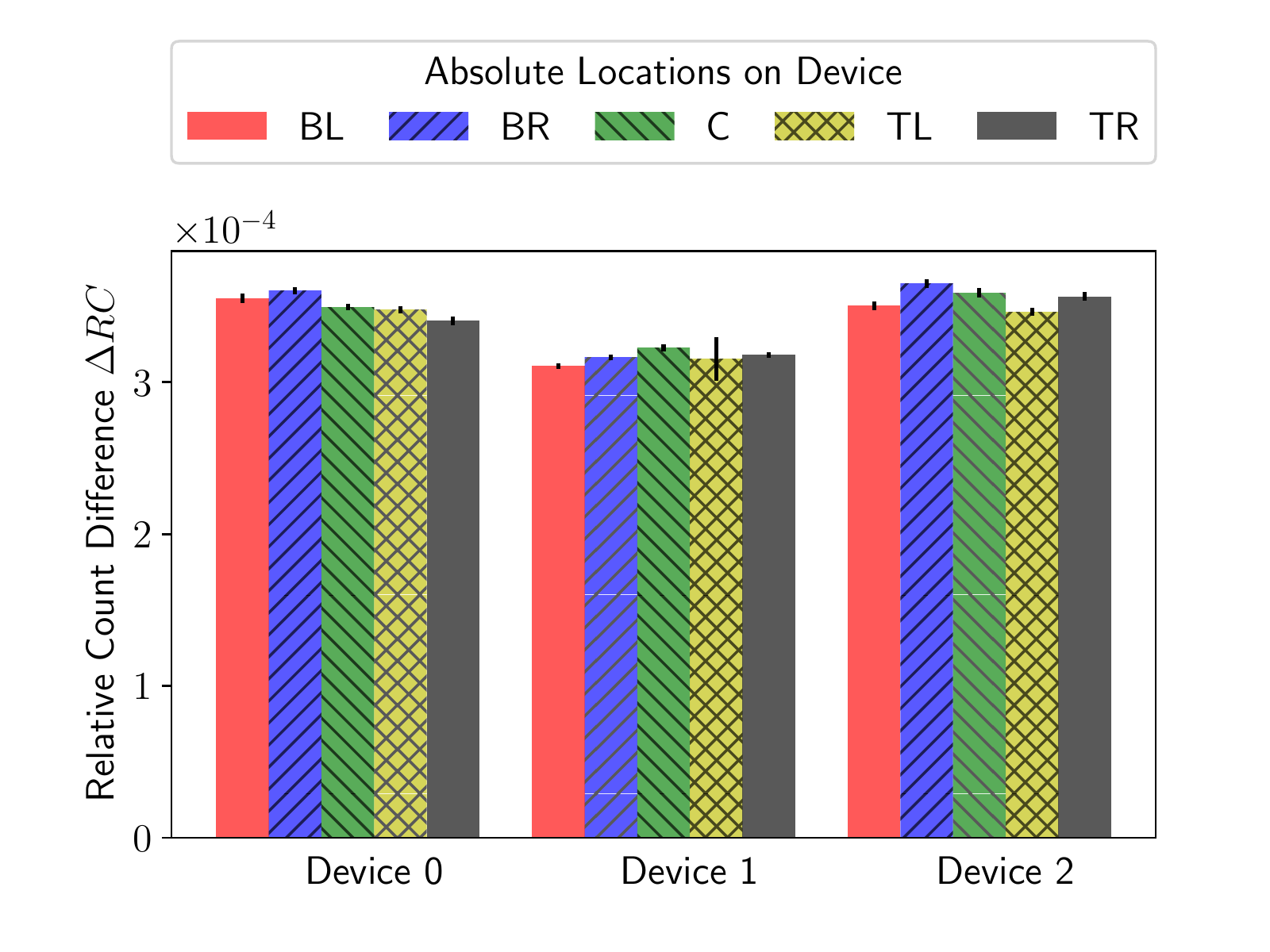}}\quad
  \subcaptionbox{Relative Offset $o_r$\label{fig:relative}}{%
    \includegraphics[width=.33\linewidth, trim=20 0 40 0, clip]{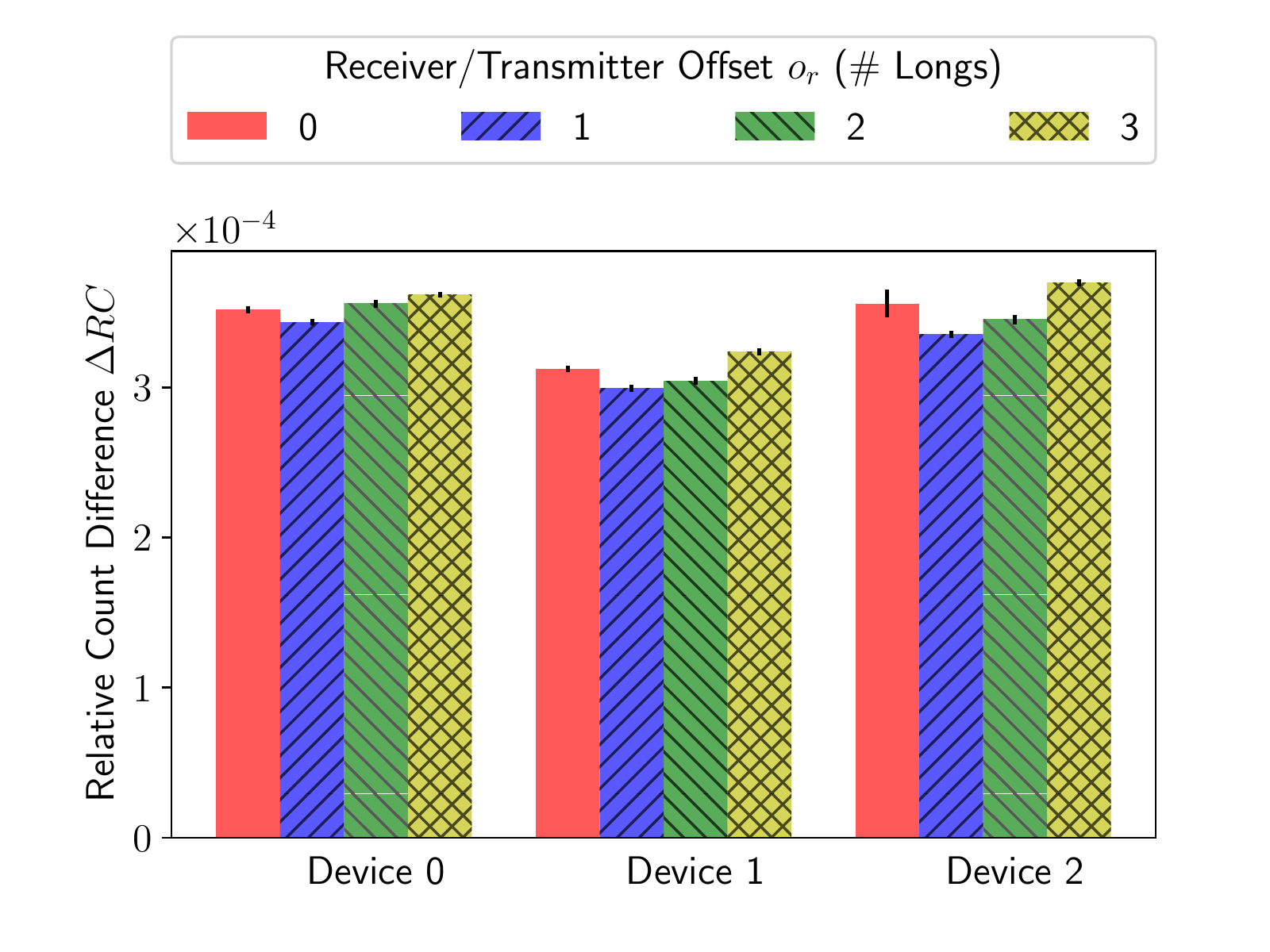}}\quad
  \subcaptionbox{Direction of Propagation\label{fig:direction}}{%
    \includegraphics[width=.33\linewidth, trim=20 0 40 0, clip]{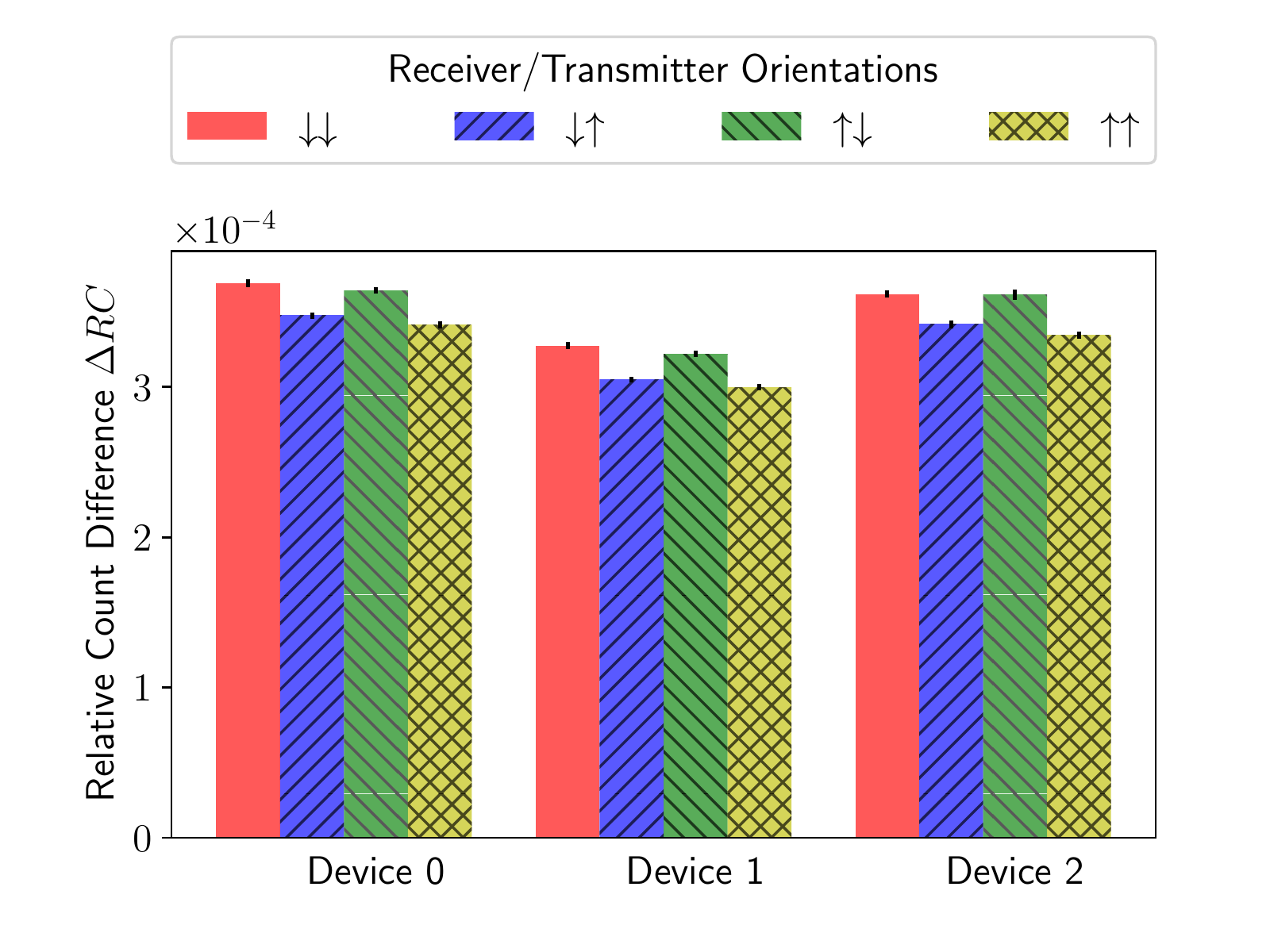}}}
\caption{Effect of location on the relative frequency of oscillation, with 99\%
confidence intervals for different placements of the circuit on the device. Absolute
location, offset, and signal orientation have little influence on the magnitude of the effect.}
\label{fig:all_locations}
\end{figure*}

\begin{figure}[tb]
  \centering
  \includegraphics[width=0.85\linewidth]{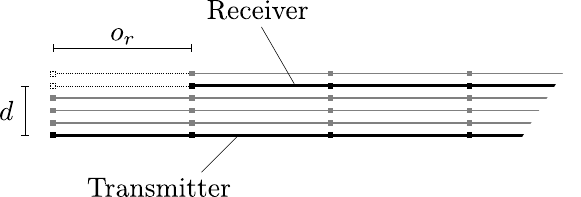}
  \caption{Relative placement of longs for transmitter and
    receiver, with respect to distance $d$ and receiver offset~$o_r$.}
  \label{fig:placement}
\end{figure}

In order to validate the location independence of the channel, we test three different
aspects of the placement of the receiver and transmitter on: the absolute location
on the device, the relative offset of the receiver and transmitter, as well as
the direction of signal propagation. Figure~\ref{fig:all_locations} shows the results
for all three experiments on the Virtex 5 devices, with 99\% confidence intervals. At a high level,
the effect remains approximately constant for each device regardless
of the choice of parameters. Across devices, the absolute magnitude of the effect
varies slightly, but is otherwise almost the same. Any variability across devices is
to be expected, since manufacturing variations
are known to affect ring oscillator frequencies~\cite{ro_jitter}.

Figure~\ref{fig:location} shows the results when an identical circuit is placed on different locations
of the device: the four corners (bottom/top left/right) and the center.
Both transmitter and receiver use 2 longs each, and they are adjacent: when the receiver's
location is $(x_r,y_r)$, the transmitter's location is $(x_t,y_t)=(x_r,y_r-1)$.
Within a device, the values are close, and there is no pattern in how the values
change between devices. Manufacturing
variations within and between devices can thus explain any variability.

The second experiment investigates the effect of the placement of the receiver and the
transmitter relative to each other. When the receiver and transmitter have different lengths,
it is possible for the two circuits to have the same overlap, but a different starting offset.
This relative offset $o_r$ (visually shown in Figure~\ref{fig:placement}) also has minimal
effect on the channel. To test this hypothesis,
we place a transmitter made up of 5 longs at a fixed location on the device.
The receiver, which uses 2 longs, is placed adjacent to the transmitter, but at an offset of $o_r$
full long wires, allowing for four different offset placements. This offset needs to correspond
to full long wire lengths due to constraints imposed by the routing architecture of the device.
Any other offset would increase the distance $d$ between the transmitter and receiver, which
we investigate separately in Section~\ref{sec:others}.
Figure~\ref{fig:relative} presents the results of this experiment,
which show approximately the same consistency both within and between
devices as those of the previous experiment.

Note that the relative effect of placing the receiver at various offsets forms a consistent
pattern across devices. As an example, the effect for an offset $o_r=3$ is consistently stronger
than it is for $o_r=1$.
This pattern can be explained by the FPGA routing layout: as mentioned in
Section~\ref{sec:background}, the local routing used to get to the various long wire segments is
different between each test. Because the local routing resources differ, the
ratio between the delay incurred by the long wire segments and the local routing resources
changes.  As will be discussed in Section~\ref{sec:others}, while the delay of the long wire
segments is affected by the transmitter, the local routing is not.

Using the same setup, and with an offset of $o_r=2$ full long wires, we change the direction of
signal propagation for the transmitter and receiver. In the previous experiments, both signals
travelled from the bottom of the device to the top. However, in the Virtex 5 architecture,
\vl{} wires are bi-directional, and can thus propagate signals upwards or downwards.
Figure~\ref{fig:direction} shows the results for the 4 different orientations
(receiver and transmitter down, receiver down/transmitter up, etc.). The
relative count difference is approximately the same for all configurations,
although as with the previous experiment, we notice a consistent ordering for
the four transmission directions across devices. Similar to the earlier experiment,
this pattern can also be explained by the routing layout.

The results of this section illustrate that only the long wires need to be
manually specified, while the registers, LUTs, and local routing can be auto-placed/routed,
further reducing the attack complexity.

\section{Resilience To Countermeasures}
\label{sec:others}

\begin{figure}[tb]
  \centering
  \includegraphics[width=.85\linewidth]{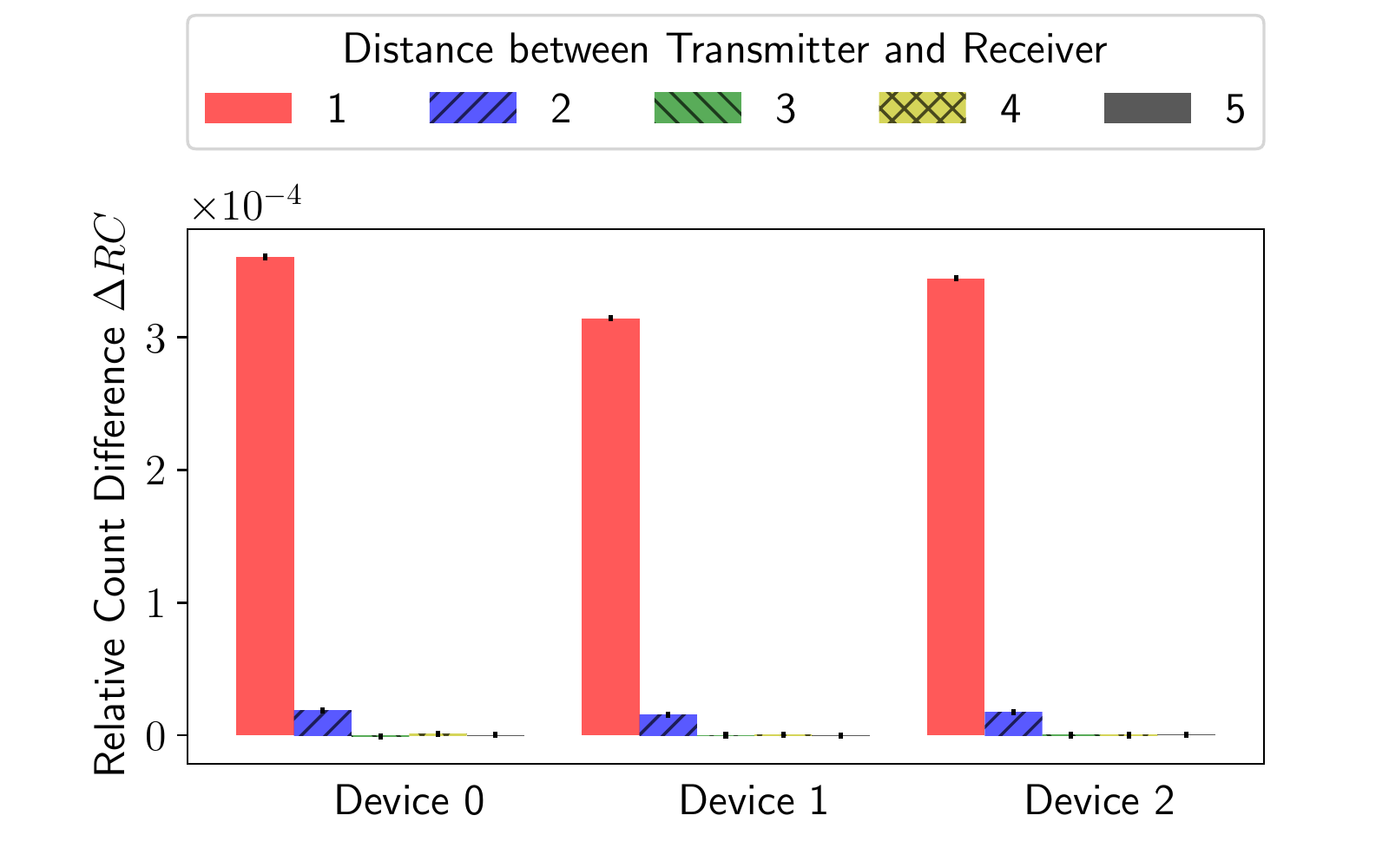}
  \caption{Effect of the transmitter-receiver distance. Long wires leak
           information up to two wires away. Distance is defined as
           in Figure~\ref{fig:placement}.}
  \label{fig:distance}
  \end{figure}

Although we discuss defense mechanisms in more depth in
Section~\ref{sec:evaluation}, in this section we evaluate
how close to the transmitter a receiver would have to be in order to decode
a message. We do this by varying the distance $d$
(depicted in Figure~\ref{fig:placement})
between the transmitter and the receiver. The results are shown in Figure~\ref{fig:distance}.
We see that the phenomenon is still measurable when separating the wires by a distance of
$d=2$, but the effect is $20$ times weaker. When the wires are farther apart ($d\ge3$),
there is no correlation between the transmitted and received values, i.e.,
the data comes from the same distribution according to the \ks{} test ($p>0.75$).
In other words, any defensive monitoring must be routed within a
distance of two to detect a transmission through the channel, and occupy
all 4 wires adjacent to a signal in order to prevent a channel from operating
successfully.

\begin{figure}[tb]
\centering
\includegraphics[width=0.8\linewidth]{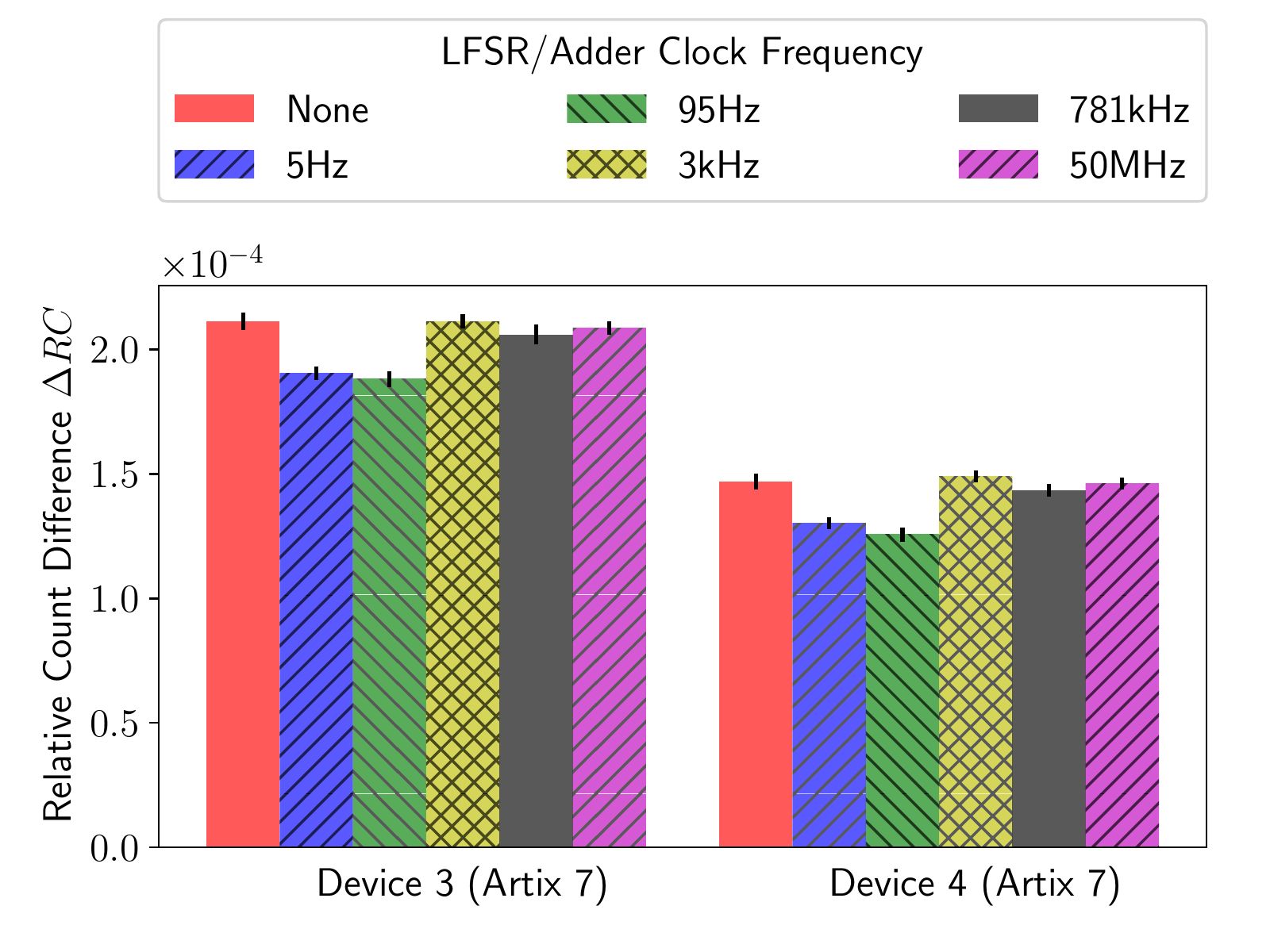}
\caption{Effect of activity induced by adders and LFSRs at different clock frequencies.
The additional activity has very little impact on channel quality.}
\label{fig:a7_adders}
\end{figure}

To test whether an active protection mechanism can disrupt the
channel through additional dynamic activity on the device, we measure the strength of the channel
in the presence of large, competing circuits which are both in- and out-of-sync
with respect to the transmissions. We synthesize 2 large 4096-bit adders,
adding different parts of a bitstream produced by a Linear Feedback Shift Register (LFSR).
As a result, both the addends and the sums change every time the LFSR produces a new bit.
The bits of each sum are then XORed together and drive 2 LEDs for
additional current draw. We run the experiment on two Artix 7 Nexys 4 boards,
for a transmitter and receiver using 10 longs each.

In order to test transmission and reception under surrounding activity of different switching
frequencies, we vary how often the LFSR produces new values by
dividing the clock driving it by $2^m$, for $m\in\{1,7,15,20,24\}$, giving us frequencies of
5Hz -- 50MHz.
The results for the two devices, including the base case of no adders and LFSRs,
are summarized in Figure~\ref{fig:a7_adders}, showing that additional
activity cannot disrupt the transmissions.
However, we note some correlation between the frequency of the activity and the
corresponding count difference. The resulting change is not sufficient
to hinder transmission, but can be used by the adversary to detect the level
of activity on the device, a technique already used by Hardware Trojan
detectors~\cite{asic_ht,ron}.

\section{Exploiting the Leakage}
\label{sec:exploiting}

In this section we discuss
exploiting the information leakage from a theoretical perspective.
In some cases (such as that of Figure~\ref{fig:v5_5}),
a threshold is sufficient for distinguishing between 0s and 1s, but
in other setups (such as that of Figure~\ref{fig:patterns}),
this separation might not be as clear:
the RO frequency may drift due to changes in environmental conditions,
such as temperature and voltage variation.
We first detail an encoding scheme
that enables high-bandwidth covert transmissions (Section~\ref{sec:classification}),
and then explain how to eavesdrop on dynamic signals through repeated
measurements (Section~\ref{sec:exfiltration}).

\subsection{Covert Transmissions}
\label{sec:classification}

To overcome the hurdle posed by local fluctuations, we propose
a Manchester encoding scheme, where
0s are transmitted as the pair $(0,1)$, and 1s as the pair $(1,0)$.
Since every pair contains each bit once,
one can decode the received pair $(c_0, c_1)$ as a 0 if $c_0<c_1$ and as a 1
otherwise. Using this scheme, transmissions
lasting $82\us$ using 2 longs as well as transmissions lasting $21\ms$ using $\frac{1}{3}$
of a long are both recovered with accuracies of 99.0 -- 99.9\%, without employing
any error correction algorithms. Under this encoding
scheme, the bandwidth of the channel is $1/(2\cdot82\cdot 10^{-6})=6.1\kbps$.

To further distinguish between noise and legitimate transmissions, we can
introduce $N$-bit start- and end-of-frame patterns. Assuming
that the probability that $c_0<c_1$ when no transmission is taking place is $1/2$ (i.e.,
each measurement is equally likely to be interpreted as a 0 or a 1), then the
probability that noise is interpreted as a start-of-frame when no transmission is taking
place is $2^{-N}$. As the channel is resilient to noise (also see Section~\ref{sec:others}),
noise will not accidentally end a transmission early or introduce additional errors while
a transmission is taking place. $N$ can thus be chosen based on the desired application guarantees,
which can include additional checksums for error detection and correction. In particular, line
codes such as 8b/10b provide single-bit error-detection capabilities, and aid in clock
recovery, making them ideal for such an application. The bandwidth of the channel is then
reduced to $6.1\times8/10\approx4.9\kbps$.

\subsection{Signal Exfiltration}
\label{sec:exfiltration}

If an adversary is merely eavesdropping on nearby signals, it is unlikely that they
will remain constant throughout the period of measurement. However, as shown in
Section~\ref{sec:dynamic} (Figure~\ref{fig:dynamic}), the delay of the long wire
depends only on the proportion of time for which the nearby
wire is carrying a 1, and {\em not} its switching frequency. This fact
reveals the Hamming Weight of the transmission during the measurement period.
By repeating measurements with a sliding window, an eavesdropping adversary
can fully recover nearby dynamic signals such as cryptographic keys with high probability.

\begin{figure}[tb]
  \centering
  \includegraphics[width=\linewidth]{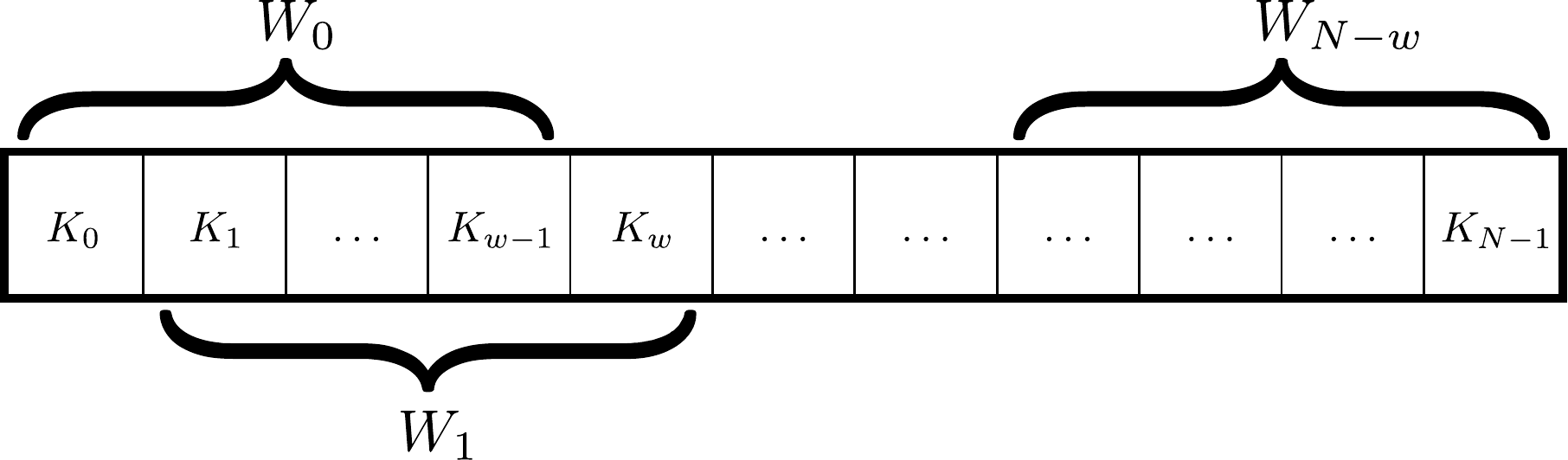}
  \caption{A sliding window of width $w$ can determine the
  relationship between key bits $K_i$ and $K_{i+w}$.}
  \label{fig:window}
\end{figure}

Suppose the adversary wishes to recover an $N$-bit key $K$, and assume
that in one period of measurement, the long wire carries $w$ consecutive bits of the key.
We assume initially that $N=nw$ is an integer multiple of the measuring window $w$, and
explain how to remove this assumption in Appendix~\ref{app:proof}.
By making repeated measurements of different but overlapping windows, as shown in
Figure~\ref{fig:window}, the adversary can recover the key with high probability.
Specifically, assume the
Hamming weight (measured by the RO count) of the first $w$ key bits $K_0$ to $K_{w-1}$
(window $W_0$) is $c_0$, and that the Hamming weight
of bits $K_1$ through $K_w$ (window $W_1$) is $c_1$. Then, if $c_0\approx c_1$ (within
some device-dependent tolerance), we can conclude that $K_0=K_{w}$.
If $c_0 > c_1$ then $K_0=1$ and $K_{w}=0$, while if $c_0 < c_1$ then $K_0=0$ and $K_{w}=1$.
By comparing the next count $c_2$ to $c_1$, one can determine the values of
$K_1$ and $K_{w+1}$, and, more generally, by repeating this process, one can determine
the relationship between $K_i$ and $K_{i+w}$.

Assuming a randomly generated key, the probability that $K_i=K_j$ for
$i\ne j$ is $1/2$. The probability that all of $S_r=(K_r$, $K_{w+r}$, \ldots, $K_{(n-1)w + r})$
are equal is $1/2^{n-1}$, since there are $n-1$ such pairs.
The probability that at least one of the bits in $S_r$ is
different than the rest is thus $1-1/2^{n-1}$. If at least one is different,
we can recover all of these bits. Repeating this argument for all possible remainders
$0\le r < w$, the probability of recovering the entire key is
\begin{align}
P=\left(1-\frac{1}{2^{n-1}}\right)^w\ge 1 - \frac{w}{2^{n-1}}
\label{eq:bound}
\end{align}
by Bernoulli's inequality. Even if it might appear counter intuitive,
the expression shows that longer keys are easier to recover than short
keys. A larger window size $w$ relative to the key length makes
recovering the key harder as there are fewer measurements over the
length of the key. For the same reason, a longer key will increase the
recovery probability. This means that asymmetric keys, e.g., those used for
signature verification are relatively easy to recover, as they are
typically much longer than symmetric keys.

\begin{figure}[tb]
  \centering
  \includegraphics[width=\linewidth]{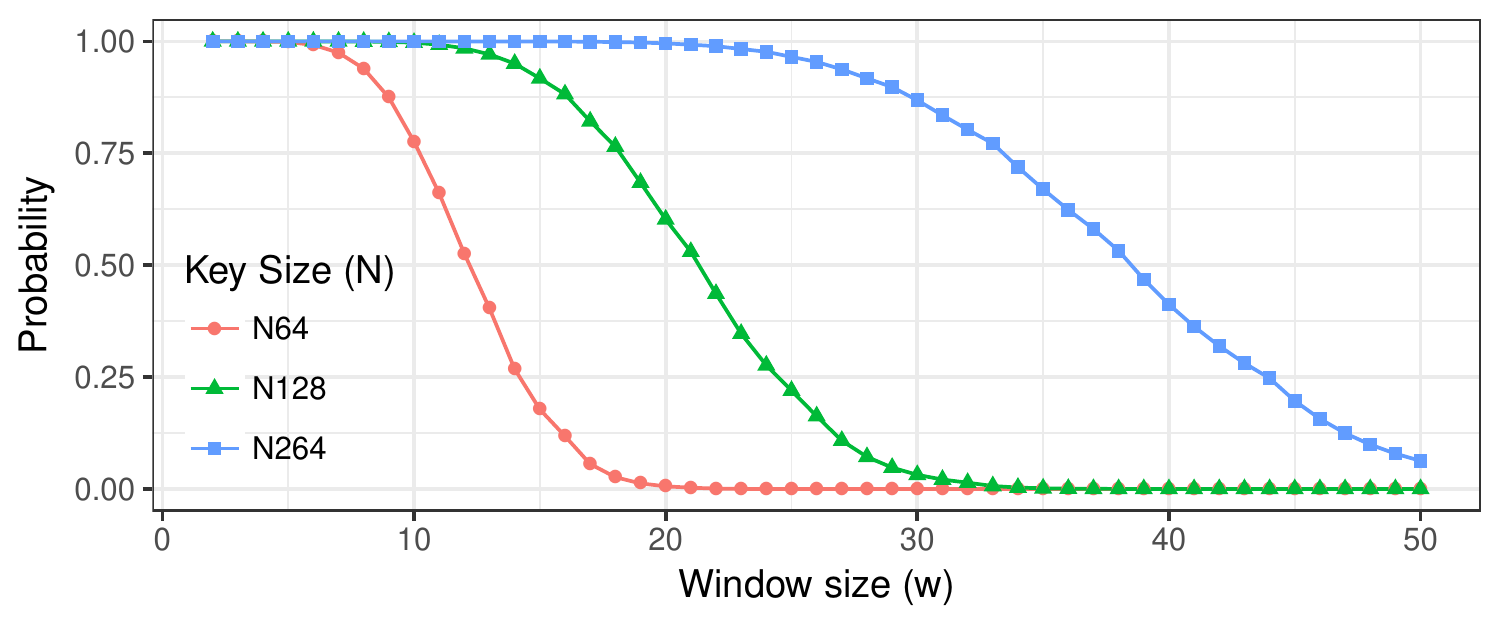}
  \caption{Probability of recovering all $N$ bits of a key based on
  the window size $w$. Smaller window sizes and longer keys yield
  a better chance to fully recover the key.}
  \label{fig:exfiltration}
\end{figure}

Figure~\ref{fig:exfiltration} shows the probability of successfully recovering \emph{all} $N$
bits of a key as a function of the window size $w$. A window of 10 bits can fully recover
a key of size 64 with 78\% probability, while a 30-bit window can recover a 264-bit key
with 87\% probability.
Figure~\ref{fig:exfiltration} only shows the probability for
full recovery, i.e., if all but one bit is recovered we still consider it a
failure. However, the above procedure can still reveal a lot about the key,
even when it does not recover it fully: in the worst
case (if the entire key consists of a repetition of its first
$w$ bits), our approach reduces the guessing space from~$2^N$ to~$2^w$ possibilities.

Extending the procedure to use multiple window lengths, we can recover any key pattern with
probability 1, except for when all key bits are the same (i.e., all ones or all
zeroes). This full recovery can be achieved using a total of just $2w+1$ full passes over the key.
Appendix~\ref{app:proof} gives an expression for the exact probability of full key recovery
for both single and multiple window sizes.

\section{Discussion}
\label{sec:evaluation}

We structure our discussion in three parts: the channel itself (Section~\ref{sec:channel}),
the cause of the information leakage (Section~\ref{sec:cause}), and potential
defenses (Section~\ref{sec:defenses}).

\subsection{The Channel}
\label{sec:channel}

The channel characterized in the previous sections
does not require any modifications to the device or special tooling, allowing an adversary
to distribute it as IP blocks. The only routing that needs to be specified
is the use of the long wires, and the only placement constraint is that the
receiver and transmitter longs are adjacent. The channel requires very little
logic: the entire setup including the signal generation
and measurement portions uses just 71 lookup tables (LUTs) and 66 registers,
excluding resources taken up by ChipScope to transfer the measurements to a PC for analysis.
As an example, our channel would only use 0.2\% of the
33,000 LUTs used in the open-source
N200/N210 Universal Software Radio Peripheral (USRP)
software-defined radio project~\cite{usrp_usage}.

The USRP source codebase~\cite{usrp_code} illustrates how IP from different organizations
makes it into a project: USRP uses code from
Ettus Research, Xilinx, Easics NV, and OpenCores (written by different authors).
Since third-party code is a necessity, and as modern IP blocks can be quite large,
the potential for unintended interaction between different cores increases.
An adversary can exploit the routing algorithms, which are forced to route
through otherwise monolithic black-box IPs due to resource constraints,
enabling his blocks to communicate covertly or eavesdrop on nearby signals.

As ring oscillators have legitimate uses,
from thermal and device health
monitors~\cite{thermal_monitoring,ro_sensing} to Hardware Trojan
detectors~\cite{asic_ht,ron}, TRNGs~\cite{trng}, and
PUFs~\cite{puf_characterization}, the adversary can make
dual-use transmitters and receivers.
The channel we identify exists whether transmissions are intentional or not,
and is a threat when an adversary controls one or more IP cores.
Unintentional transmissions pose new risks for multi-user
scenarios, including FPGA/CPU hybrids and cloud infrastructures
offering FPGA solutions. In these setups, an adversarial receiver can
be placed next to long wires used by other third-party vendors
and eavesdrop on the signals carried by them. The same mechanism can also
be used for legitimate purposes such as watermarks
and no-contact debugging taps.

\subsection{Leakage Cause}
\label{sec:cause}

So far, we have focused on the novelty and applicability of the phenomenon
presented, rather than its cause. In Section~\ref{sec:local}, we showed
that the phenomenon depends on the use of the long wires, and not the
switching activity of circuits, which decreases rather than increases
ring oscillator frequency. The only other work which deals with
long wires delays is~\cite{ro_crosstalk}, where a RO with a long wire
was placed next to other long wires carrying signals which where
either equal to the RO signal or opposing it. It was shown that
when a nearby long wire has the same value as the RO wire, the
frequency of the RO is higher compared to the RO frequency when
the nearby long wire has the opposite value (i.e., if the current
value on the RO long wire is a 1, the value on the nearby wire
is 0 and vice versa). The work in~\cite{ro_crosstalk} necessitates that
the signal of the RO and the nearby wire be in sync, so the
wires were directly connected, and static patterns which are
independent of the RO signal were not tested. By contrast,
in our work, we showed that nearby wires are influenced even when there is no
connection between the transmitter and the receiver, and even when
the transmitted value remains constant during the measurement period.
These two properties can be
exploited in constructing a communication channel.

Although~\cite{ro_crosstalk} broadly categorized their observations as ``capacitive crosstalk'',
it made no attempt to precisely determine the physical cause behind it. This would indeed
be difficult without design information such as
physical layout and process-specific parameters. This ``lack of electrical detail''
on FPGAs is, in fact, well-known and has been identified by multiple
authors~\cite{bist_crosstalk,interconnect_capacitance,crosstalk_noise,builtin_crosstalk,crosstalk_router}.

As a result, whether the effect we have found exists due to drive-strength issues,
electromagnetic emissions, or some other property of FPGAs remains an open question.
It is even possible that the wires themselves might not be the cause of the issue,
but that the buffers driving them share local connections to the power network. However,
without more specialized equipment to x-ray the chips to further narrow
down the potential causes, we cannot determine the precise cause, or even whether
ASICs would be affected.

Overall, the characterization of the channel is valuable
even without access to these details, since we have shown it to always be present
and easily measurable on off-the-shelf devices without special
modifications. FPGA users cannot alter the electrical behavior of the device,
but can only influence how circuits are mapped onto the FPGA.
As a result, FPGA circuit designers cannot change the existence of the channel,
and need to be aware of the communication and exfiltration capabilities that this channel
introduces.

\subsection{Defense Mechanisms}
\label{sec:defenses}

Section~\ref{sec:others} showed that
one cannot detect transmissions from a distance $d\ge2$,
and that spurious activity (in the form of adders and additional current draw)
does not eliminate the transmission channel. Hence,
defense mechanisms need to protect a design before it is loaded onto the FPGA.
Since long wires are an integral part of the reconfigurable FPGA fabric,
detecting the transmitter is not easy: the long can
be used as part of the connections within an IP block, carrying sensitive information.
Routing algorithms thus need to be modified to account for this information leakage,
by introducing directives which mark signals, or even entire blocks as sensitive.
The tools then need to add ``guard wires'', by either leaving the four nearby
long wires unoccupied, or by occupying them with compiler-generated random signals.
We note that even though this approach will prevent the leakage from occurring,
it is particularly taxing for dense designs, and can make placement and routing
more time-consuming, or even lead to timing violations.

Designers using unpatched tools need to be aware of this source of leakage,
and must either manually look for long-wires post-routing, explicitly
add guard wires, or, more generally, specify placement and routing constraints
for both highly-sensitive signals, and untrusted third-party blocks.
Overall, better defense mechanisms for future FPGA generations are needed
at the architectural level,
and require a deeper understanding of the cause of this phenomenon.

\section{Related Work}
\label{sec:related}

Research on side- and covert-channels on FPGAs and other embedded devices has primarily
focused on communications between the device and the outside world.
Techniques include varying the power consumption
of a device and measuring the impulse response~\cite{power_side}, changing how much
Electromagnetic Interference is emitted by the device~\cite{emi_covert},
or, in the other direction of communication, measuring voltage~\cite{voltage_side}
and temperature changes~\cite{side_receivers}.
These side-channels can be employed in the context of creating Hardware Trojans
(HTs)~\cite{trojan_side}, or as ways to watermark circuits and offer IP
protection~\cite{side_watermarks,voltage_side}.

Many of these circuits employ ring oscillators, exploiting their dependence
on Process, Voltage, and Temperature (PVT) variations~\cite{ro_jitter}.
ROs are primarily used on the receiving end, but they can also be used
to transmit information by causing changes in temperature~\cite{temperature_covert}.
This technique allows communications between blocks on the same device,
under a threat model similar to ours.

Ring oscillators have also been used in security-sensitive applications, including
True Random Number Generators (TRNGs)~\cite{trng} and
Physically Unclonable Functions (PUFs)~\cite{puf_characterization}.
Consequently, any mechanism which can be used to manipulate or bias their frequency
can also be used to attack these applications. Besides the technique
introduced in this paper, prior work has
influenced the delays of ROs by altering the
power supply~\cite{trng_injection} and by injecting EM signals~\cite{trng_em},
resulting in low entropy and cloneability.

As explained in Section~\ref{sec:cause}, a switching pattern in sync with the
RO's signal increases the RO's oscillation
frequency by 1-9\% compared to a pattern that opposes it~\cite{ro_crosstalk}.
To achieve this synchronization, however, requires the transmitter to
be connected to the output
of one of the RO's stages. As a result, as presented, this
mechanism cannot be used directly for side-channel
communication or to reliably attack the ring oscillator, due to
the high accuracy of prediction required for the frequency and
phase of the oscillator.

Emphasis has also been placed on using networks of ROs
to detect Hardware Trojans on a device~\cite{asic_ht,ron}.
The dynamic power consumed by HTs results in a voltage drop that lowers the RO
frequencies compared to those in the Trojan-free ``golden'' IC, making them detectable.
Such prior work depends on the effect switching activity has on the frequency of ROs to detect
HTs. As shown in Figure~\ref{fig:dynamic_local}, when using short wires, we were able to
reproduce the prior effect, where only the number of bit transitions, and not the actual bits
themselves, were the cause for the RO frequency drop. However, this no longer holds for
long wires: the frequency increases based on the duration for which a 1 is transmitted,
irrespective of the dynamic activity. As a result, the channel depends on a fundamentally different
phenomenon, which uses the values carried on the wires themselves, and not their transitions.

Prior work~\cite{static_leakage} has shown that to detect slowly-changing signals,
very large circuits (over 14k registers) or long measurement times (2.5h) are needed,
in addition to external measurement equipment,
and special modifications to the device.
By contrast, our work only uses small on-chip circuits, without any special
control over voltage or temperature conditions. We can distinguish
between the values of signals which
remain constant (i.e., have no switching activity) during our period of measurement,
which is as low as $82\us$, a measurement period which is also a lower bound
for on-chip HT detection using ROs~\cite{ro_ht_analysis}.

\section{Conclusion}
\label{sec:conclusion}

We demonstrated the existence of a previously
unexplored phenomenon on FPGA devices, that causes the delay of so called ``long
wires'' to depend on the logical state of nearby long wires,
{\em even when the driven value remains constant}. The effect
is small, but surprisingly resilient, and measurable within the device
by small circuits even in the presence of environmental noise,
and without any modifications to the FPGA.  We use
this phenomenon to create a communication channel between circuits that are
not physically connected.  As designs often incorporate circuits from
multiple third-parties, this channel can break
separation of privilege between IP cores of different
trust levels, or enable communication between distinct
cores in multi-user setups. Such use-cases are increasingly
common as FPGAs and CPUs become integrated, and as FPGAs become
available on public cloud infrastructures. The same mechanism can also be used
to eavesdrop and recover keys with high probability even when the signals
change during the period of measurement, or to implement a no-contact
debugging tap or watermark scheme.
In our prototype implementation, the channel has a bandwidth of up to $6\kbps$,
and we can recover over 99\% of the transmitted bits
correctly using a Manchester encoding scheme.
We showed that the phenomenon is present in three generations of Xilinx FPGAs,
and that the channel can be implemented in a
variety of arrangements, including different locations, orientations,
and with multiple
transmitting circuits present.  The strength of the phenomenon scales
linearly with the number of wires used, and also dominates a
competing effect caused by switching activity.

\appendix

\section{Generalizing Signal Exfiltration}
\label{app:proof}

\balance

In this section we explain how to remove the assumption that
the key size $N$ is a multiple of the window size $w$, and how
to fully recover keys by varying the window size.

To start with, if $N=nw+m$, with $0\le m<w$, the probability that the
bits in $S_r=\big(K_r$, $K_{w+r}$, $K_{2w + r}$, $\ldots\big)$
are the same is $1/2^{n}$ for $0\le r < m$, since $|S_r|=n+1$.
For $m \le r < w$ this probability is $1/2^{n-1}$, as $|S_r|=n$. This allows
us to adjust Equation (\ref{eq:bound}) to
\begin{align}
\label{eq:bound_fixed}
P=\left(1-\frac{1}{2^{n}}\right)^{m}\left(1-\frac{1}{2^{n-1}}\right)^{w-m}
\end{align}
In particular, if $N$ is a multiple of $w$, then $m=0$, so the
above expression reduces to~(\ref{eq:bound}).

The expression is valid for any length $N\ge 2w-1$,
removing the requirement that $N$ is an
integer multiple of $w$. The lower bound on $N$ is necessary if we wish to
recover the first $w$ bits of the key, as we need
to have $r + w \le N$ for each $r$ with $0\le r \le w - 1$ in order to have elements in $S_r$.

Suppose that the original measurements were not able to recover
the bits in $S_r$ because they were all identical. By repeating measurements
with a window of size $w+1$, the algorithm either recovers all bits in the
sequence $S'_r=\big(K_r$, $K_{w+1+r}$, $K_{2(w+1)+r}$, $\ldots\big)$
or shows that they too are identical (here, under the assumption that $N\ge 2w+1$).

In the first case, the algorithm recovers $K_r$, and hence $S_r$ since all its bits
are identical. In the second case, where all bits in $S'_r$ are also identical,
the entire key consists of a single repeated bit (i.e., all ones or all zeroes). This is because
$K_r=K_{w+1+r}=K_{r+1\pmod w}$, and $K_r=K_{2(w+1)+r}=K_{r+2\pmod w}$, etc. Note that
the size of $S'_r$ might be too small to cover the all the residues $\bmod w$ by
itself, but varying $r$ allows us to recover all of $K_0,\ldots,K_{w-1}$ with
probability~1 if there are at least 2 different bits in the key, or to determine
that the key consists of the same repeated bit.

For a window of size $w$, we need $N-w+1$ measurements, but this can be accomplished
in only $w$ independent runs of the experiment. Run $r$ is responsible for collecting the
measurements for the parts of the key that start with $K_{r+w\cdot i}$ for some $i$. For example,
run $1$ measures the Hamming weight of the key bits $\big(K_1,\ldots,K_w\big)$, and
$\big(K_{w+1},\ldots,K_{2w}\big)$, etc. As there is no overlap in the bits used, these
measurements can be completed in a single run. Thus, using both window sizes, and
to fully determine all the bits of a key, one needs to take
\[\left(N-w+1\right)+\left(N-(w+1)+1\right)=2N-2w+1\]
measurements over just
\[w+(w+1)=2w+1\]
runs, showing that this key-recovery algorithm is efficient. In other words, the key
only needs to be repeated $2w+1$ times to be fully leaked.

\end{document}